\documentclass[journal]{IEEEtran}

\usepackage{graphicx, comment}
\usepackage{tabularx}
\usepackage{booktabs}
\usepackage{multirow}
\usepackage{pmat}
\usepackage{amsmath, amsfonts}

\def\BibTeX{{\rm B\kern-.05em{\sc i\kern-.025em b}\kern-.08em
    T\kern-.1667em\lower.7ex\hbox{E}\kern-.125emX}}

\newtheorem{theorem}{Theorem}
\newtheorem{proposition}{Proposition}
\newtheorem{definition}{Definition}
\newtheorem{lemma}{Lemma}
\newtheorem{corollary}{Corollary}

\newenvironment{thmproof}[1]
{\noindent\hspace{2em}{\it #1 }}
{\hspace*{\fill}~\QED\par\endtrivlist\unskip}

\def\twobyone#1#2{\left(\begin{array}{c}#1 \\ #2\end{array}\right)}

\def\Rank{\mathop{\rm Rank}\limits}
\def\Zmodm{{\mathbb{Z}}_m}
\def\GF{\mathop{\sf GF}\limits}
\def\bigorder{{\mathcal{O}}}
\def\RR{{\mathbb{R}}}
\def\NN{{\mathbb{N}}}
\def\PP{{\mathsf{P}}}
\def\EE{{\mathsf{E}}}
\def\const{{\mathsf{const}}}

\newcounter{mytempeqncnt}

\def \Definition#1{{\it Definition~#1}}

\def \Theorem#1{{\it Theorem~#1}}
\def \Theorems#1{{\it Theorems~#1}}
\def \Corollary#1{{\it Corollary~#1}}

\def \Proposition#1{{\it Proposition~#1}}

\def \Lemma#1{{\it Lemma~#1}}

\def \Figure#1{{\rm Fig.~#1}}
\def \Figures#1{{\rm Figs.~#1}}
\def \Table#1{{\rm TABLE~#1}}

\def \Append#1{{\sc Appendix~#1}}

\def\picchk{\put(0,0){\circle{30}}\put(0,0){\makebox(0,0){\small$+$}}}

\def\cmpt{\hspace{-.015cm}}

\begin{document}
%
%
\title{Density Evolution for Asymmetric Memoryless Channels}

\author{Chih-Chun~Wang, Sanjeev~R.~Kulkarni,~\IEEEmembership{Fellow,~IEEE,}
H.~Vincent~Poor,~\IEEEmembership{Fellow,~IEEE,}\thanks{Manuscript received
August 15, 2002; revised August 22, 2005.
        This work was supported in part by  the National Science Foundation under
        Grants
No.~CCR-9980590 and CCR-0312413, the Army Research Laboratory Communications
Technology Alliance under Contract No.~DAAD19-01-2-0011, the Army Research
Office under Contract No.~DAAD19-00-1-0466, and the New Jersey Center for
Pervasive Information Technologies.}
\thanks{This
paper was presented in part at the 3rd International Symposium on Turbo Codes
\& Related Topics, Brest, France, Sept. 1--5, 2003, and in part at the 39th
Annual Conference on Information Sciences and Systems,  Baltimore, USA, March
16--18, 2005.}\thanks{The authors are with the Department of Electrical
Engineering, Princeton University, Princeton, NJ 08544. Email: \{chihw,
kulkarni, poor\}@princeton.edu} } \maketitle

\markboth{IEEE Transactions on Information Theory,~Vol.~51,
No.~8,~August~2005}{Wang \MakeLowercase{\textit{et al.}}: Density Evolution for
Asymmetric Memoryless Channels}


\begin{abstract}
Density evolution is one of the most powerful analytical tools for low-density
parity-check (LDPC) codes and graph codes with message passing decoding
algorithms. With channel symmetry as one of its fundamental assumptions,
density evolution (DE) has been widely and successfully applied to different
channels, including binary erasure channels, binary symmetric channels, binary
additive white Gaussian noise channels, etc. This paper generalizes density
evolution for {\it non-symmetric} memoryless channels, which in turn broadens
the applications to general memoryless channels, e.g.\ z-channels, composite
white Gaussian noise channels, etc. The central theorem underpinning this
generalization is the convergence to perfect projection for any fixed size
supporting tree. A new iterative formula of the same complexity is then
presented and the necessary theorems for the performance concentration theorems
are developed. Several properties of the new density evolution method are
explored, including stability results for general asymmetric memoryless
channels. Simulations, code optimizations, and possible new applications
suggested by this new density evolution method are also provided. This result
is also used to prove the typicality of linear LDPC codes among the coset code
ensemble when the minimum check node degree is sufficiently large. It is shown
that the convergence to perfect projection is essential to the belief
propagation algorithm even when only symmetric channels are considered. Hence
the proof of the convergence to perfect projection serves also as a completion
of the theory of classical density evolution for symmetric memoryless channels.
\end{abstract}


\section{Introduction}
\PARstart{S}{ince} the advent of turbo codes
\cite{BerrouGlavieux96} and the rediscovery of low-density
parity-check (LDPC) codes \cite{Gallager63,MacKay99} in the mid
1990's, graph codes \cite{SipserSpielman96} have attracted
significant attention because of their capacity-approaching error
correcting capability and the inherent low-complexity
($\bigorder(n)$ or $\bigorder(n\log(n))$ where $n$ is the codeword
length) of message passing decoding algorithms \cite{MacKay99}.
The near-optimal performance of graph codes is generally based on
pseudo-random interconnections and Pearl's belief propagation (BP)
algorithm \cite{Pearl88}, which is a distributed message-passing
algorithm efficiently computing {\it a posteriori} probabilities
in cycle-free inference networks. Turbo codes can also be viewed
as a variation of LDPC codes, as discussed in \cite{MacKay99} and
\cite{McElieceMacKayCheng98}.

Due to their simple arithmetic structure, completely parallel decoding
algorithms, excellent error correcting capability
\cite{ChungForneyRichardsonUrbanke01}, and acceptable encoding complexity
\cite{RichardsonUrbanke01b,Spielman96}, LDPC codes have
 been widely and successfully applied to different channels, including binary erasure channels
(BECs)
\cite{ByersLubyMitzenmacher02,LubyMitzenmacherShokrollahiSpielman98,LubyMitzenmacherShokrollahiSpielman01},
binary symmetric channels (BSCs), binary-input additive white
Gaussian noise channels (BiAWGNCs)
\cite{MacKay99,RichardsonUrbanke01a}, Rayleigh fading channels
\cite{HouSiegelMilistein01}, Markov channels \cite{GarciaFrias01},
partial response channels/intersymbol interference channels
\cite{KavcicMaMitzanmacher03,KurkoskiSiegelWolf02,LiNarayananKurtasGeorghiades02,ThangarajMcLaughlin02},
dirty paper coding \cite{CaireBurshteinShamai02}, and
bit-interleaved coded modulation
\cite{HouSiegelMilisteinPfister03}. Except for the finite-length
analysis of LDPC codes over the BEC
\cite{DiProiettiTelatarRichardsonUrbanke02}, the analysis of
iterative message-passing decoding algorithms is
    asymptotic (when the block length tends to infinity) \cite{RichardsonUrbanke01a,RichardsonShokrollahiUrbanke01}.
Under the optimal maximum-likelihood (ML) decoding algorithm, both the
finite-length analysis and the asymptotic analysis for LDPC codes and other
ensembles of turbo-like codes become tractable and rely on the weight
distribution of these ensembles (see e.g.
\cite{LitsynShevelev02,JinMcEliece02}, and \cite{Lehmann03}). Various Gallager
type bounds on ML decoders for different finite LDPC code ensembles have been
established in \cite{ShamaiSason02}.

In essence, the density evolution method proposed by Richardson
{\it et al.}\ in \cite{RichardsonUrbanke01a} is an asymptotic
analytical tool for LDPC codes. As the codeword length tends to
infinity, the random codebook will be more and more likely to be
cycle-free, under which condition the input messages of each node
are independent. Therefore the probability density of messages
passed can be computed iteratively. A performance concentration
theorem and a cycle-free convergence theorem, providing the
theoretical foundation of the density evolution method, are proved
in \cite{RichardsonUrbanke01a}. The behavior of codes with block
length $>10^4$ is well predicted by this technique, and thus
degree optimization for LDPC codes becomes tractable. Near optimal
LDPC codes have been found in
\cite{ChungForneyRichardsonUrbanke01} and
\cite{RichardsonShokrollahiUrbanke01}. In
\cite{KavcicMaMitzanmacher03} Kav\v{c}i\'{c} {\it et al.}\
generalized the density evolution method to intersymbol
interference channels, by introducing the ensemble of {\it coset
codes}, i.e.\ the parity check equations are {\it randomly}
selected as even or odd parities. Kav\v{c}i\'{c} {\it et al.}\
also proved the corresponding fundamental theorems for the new
coset code ensemble.

Because of the symmetry of the BP algorithm and the symmetry of
parity check constraints in LDPC codes, the decoding error
probability will be independent of the transmitted codeword in the
symmetric channel setting. Thus, in \cite{RichardsonUrbanke01a},
an all-zero transmitted codeword is assumed and the probability
density of the messages passed depends only on the noise
distribution. Nevertheless, in symbol-dependent asymmetric
channels, which are the subject of this paper, the noise
distribution is codeword-dependent, and thus some codewords are
more noise-resistant than others. As a result, the all-zero
codeword cannot be assumed. Instead of using a larger coset code
ensemble as in \cite{KavcicMaMitzanmacher03}, we circumvent this
problem by averaging over all valid codewords, which is
straightforward and has practical interpretations as the averaged
error probability. Our results apply to all binary input,
memoryless, symbol-dependent channels (e.g., z-channels, binary
asymmetric channels (BASCs), composite binary-input white Gaussian
channels (composite BiAWGNCs), etc.) and can be generalized to
LDPC codes over $\GF(q)$ or $\Zmodm$
\cite{BennatanBurshtein03a,BennatanBurshtein04,WangKulkarniPoor04}.
The theorem of convergence to {\it perfect projection} is provided
to justify this codeword-averaged approach in conjunction with the
existing theorems.
  New
results on monotonicity, symmetry, stability (a necessary and a sufficient
condition), and convergence rate analysis of the codeword-averaged density
evolution method are also provided. Our approach based on the
linear\footnote{LDPC codes are, by definition, linear codes since only even
parity check equations are considered. Nonetheless, by taking both even and odd
parity check equations into consideration, the extended LDPC ``coset" code has
been proven to have important practical and theoretical value in many
applications \cite{KavcicMaMitzanmacher03}. To be explicit on whether only even
parity-check equations are considered or an extended set of parity-check
equations is involved, two terms, ``linear LDPC codes" and ``LDPC coset codes,"
will be used whenever a comparison is made, even though the adjective, linear,
is redundant for traditional LDPC codes.} code ensemble will be linked to that
of the coset code ensemble \cite{KavcicMaMitzanmacher03} by proving the
typicality of linear\footnotemark[\value{footnote}] LDPC codes when the minimum
check node degree is sufficiently large, which was first conjectured in
\cite{HouSiegelMilisteinPfister03}. All of the above generalizations are based
on the convergence to perfect projection, which will serve also as a
theoretical foundation for the belief propagation algorithms even when only
symmetric channels are considered.


This paper is organized as follows. The formulations of and
background on channel models, LDPC code ensembles, the belief
propagation algorithm, and density evolution, are provided in
Section~\ref{sec:formulations}. In
Section~\ref{sec:iterative-formula}, an iterative formula is
developed for computing the evolution of the codeword-averaged
probability density. In Section~\ref{sec:theorems}, we state the
theorem of convergence to perfect projection, which justifies the
iterative formula. A detailed proof will be given in
\Append{\ref{app:proof-convergence}}. Monotonicity, symmetry, and
stability theorems are stated and proved in
Section~\ref{sec:related-theorems}.
 Section~\ref{sec:simulations} consists of simulations and
discussion of possible applications of our new density evolution method.
Section~\ref{sec:side-results} proves the typicality of linear LDPC codes and
revisits belief propagation for symmetric channels.
Section~\ref{sec:conclusions} concludes the paper.

%
%
\section{Formulations\label{sec:formulations}}
%
%
\subsection{Symbol-dependent Non-symmetric Channels}
The memoryless, symbol-dependent channels we consider here are modeled as
follows. Let $\mathbf x$ and $\mathbf y$ denote a transmitted codeword vector
and a received signal vector of codeword length $n$, where $x_i$ and $y_i$ are
the $i$-th transmitted symbol and received signal, respectively,  taking values
in $\GF(2)$ and the reals, respectively. The channel is memoryless and is
specified by the conditional probability density function $f_{\mathbf y|\mathbf
x}({\mathbf y}|{\mathbf x})=\prod_{i=1}^nf(y_i|x_i)$. Two common examples are
as follows.

\begin{itemize}
\item {\it Example 1:} [Binary Asymmetric Channels (BASC)]
    \begin{eqnarray}
    f(y|x)=\begin{cases}
     (1-\epsilon_0)\delta(y)+\epsilon_0\delta(y-1) & \text{if $x=0$}\\
    \epsilon_1\delta(y)+(1-\epsilon_1)\delta(y-1) & \text{if $x=1$}
 \end{cases}
,\nonumber
    \end{eqnarray}
    where $\epsilon_0,\epsilon_1$ are the crossover probabilities and
    $\delta(y)$ is the Dirac delta function. Note: if
    $\epsilon_0=0$, the above collapses to the z-channel.
\item {\it Example 2:} [Composite BiAWGNCs]
{\footnotesize
    \begin{eqnarray}
    f(y|x)=\begin{cases}\frac{1}{2}
    \frac{1}{\sqrt{2\pi\sigma^2}}\left(e^{-\frac{(y-3/\sqrt{5})^2}{2\sigma^2}}+e^{-\frac{(y+3/\sqrt{5})^2}{2\sigma^2}}\right) & \text{if $x=0$}\\
\frac{1}{2}
    \frac{1}{\sqrt{2\pi\sigma^2}}\left(e^{-\frac{(y-1/\sqrt{5})^2}{2\sigma^2}}+e^{-\frac{(y+1/\sqrt{5})^2}{2\sigma^2}}\right) & \text{if $x=1$}
 \end{cases}
,\nonumber
    \end{eqnarray}}\cmpt
    which corresponds to a bit-level sub-channel of the 4~pulse amplitude modulation (4PAM) with Gray mapping.
\end{itemize}

%
%
\subsection{Linear LDPC Code Ensembles}
The linear LDPC codes of length $n$ are actually a special family of parity
check codes, such that all codewords can be specified by the following even
parity check equation in $\GF(2)$:
    \begin{eqnarray}
    {\mathbf A}{\mathbf x}={\mathbf 0},\nonumber
    \end{eqnarray}
where ${\mathbf A}$ is an $m \times n$ sparse matrix in $\GF(2)$ with the
number of non-zero elements linearly proportional to $n$. To facilitate our
analysis, we use a code ensemble rather than
 a fixed code. Our linear code ensemble is generated by equiprobable
edge permutations in a regular bipartite graph.

\begin{figure}[t]
\begin{tabular}{cc}
\parbox[c]{4cm}
{

\setlength{\unitlength}{.08mm}
    \begin{picture}(500,250)
    \put(10,12){$i=$}
    \put(82, 80){
    \put(0,-18){\circle{36}}
    \put(-12,-68){1}
    }
    \put(162, 80){
    \put(0,-18){\circle{36}}
    \put(-12,-68){2}
    }
    \put(242, 80){
    \put(0,-18){\circle{36}}
    \put(-12,-68){3}
    }
    \put(322, 80){
    \put(0,-18){\circle{36}}
    \put(-12,-68){4}
    }
    \put(402, 80){
    \put(0,-18){\circle{36}}
    \put(-12,-68){5}
    }
    \put(482, 80){
    \put(0,-18){\circle{36}}
    \put(-12,-68){6}
    }
%
    \put(162, 160){
    \put(-16,0){\framebox(32,32){}}
    \put(-12,42){1}
    }
%
    \put(242, 160){
    \put(-16,0){\framebox(32,32){}}
    \put(-12,42){2}
    }
%
    \put(322, 160){
    \put(-16,0){\framebox(32,32){}}
    \put(-12,42){3}
    }
%
    \put(402, 160){
    \put(-16,0){\framebox(32,32){}}
    \put(-12,42){4}
    }
\put(92, 202){$j=$}
%
    \put(82,80){\line(1,1){80}}
    \put(162,80){\line(0,1){80}}
    \put(242,80){\line(-1,1){80}}
    \put(325,80){\line(-1,1){80}}
    \put(319,80){\line(-1,1){80}}
    \put(482,80){\line(-3,1){240}}
    \put(162,80){\line(2,1){160}}
    \put(242,80){\line(1,1){80}}
    \put(402,80){\line(-1,1){80}}
    \put(82,80){\line(4,1){320}}
    \put(402,80){\line(0,1){80}}
    \put(482,80){\line(-1,1){80}}
\end{picture}} &
{\footnotesize  ${\mathbf A}=\left(\begin{array}{cccccc}
    1&1&1&0&0&0\\
    0&0&0&0&0&1\\
    0&1&1&0&1&0\\
    1&0&0&0&1&1
    \end{array}\right)$}
\end{tabular}
\caption{A realization of the  code ensemble ${\mathcal
C}^6(2,3)$.} \label{fig:codebook}
\end{figure}
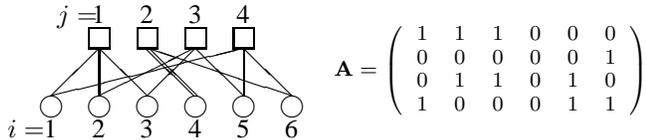

As illustrated in \Figure{\ref{fig:codebook}}, the bipartite graph model
consists of a bottom row of variable nodes (corresponding to codeword bits) and
a top row of check nodes (corresponding to parity check equations). Suppose we
have $n$ variable nodes on the bottom and each of them has $d_v$ sockets. There
are $m:=\frac{nd_v}{d_c}$ check nodes on the top and each of them has $d_c$
sockets. With these fixed $(n+m)$ nodes, there are a total of $(nd_v)!$
possible configurations obtained by connecting these $nd_v=md_c$ sockets on
each side, assuming all sockets are distinguishable.\footnote{When assuming all
variable/check node sockets are indistinguishable, the number of configurations
can be upper bounded by $\frac{(nd_v)!}{(d_c!)^m}$.} The resulting graphs
(multigraphs) will be regular and bipartite with degrees denoted by
$(d_v,d_c)$, and can be mapped to parity check codes with the convention that
the variable bit~$v$ is involved in parity check equation $c$ if and only if
the variable node~$v$ and the check node~$c$ are connected by an odd number of
edges. We consider a regular code ensemble ${\mathcal C}^n(d_v,d_c)$ putting
equal probability on each of the possible configurations of the regular
bipartite graphs described above. One realization of the codebook ensemble
${\mathcal C}^6(2,3)$ is shown in \Figure{\ref{fig:codebook}}. For practical
interest, we assume $d_c>2$.

 For each graph in
${\mathcal C}^n(d_v,d_c)$, the parity check matrix $\mathbf A$ is an $m \times
n$ matrix over $GF(2)$, with $A_{j,i}=1$ if and only if there is an {\it odd}
number of edges between variable node $i$ and check node $j$. Any valid
codeword $\mathbf x$ satisfies the parity check equation ${\mathbf A}{\mathbf
x}={\mathbf 0}$. For future use, we let $i$ and $j$ denote the indices of the
$i$-th variable node and the $j$-th check node. $\{j_{i_0,c}\}_{c\in[1,d_v]}$
denotes all check nodes connecting to variable node $i_0$ and similarly with
$\{i_{j_0,v}\}_{v\in[1,d_c]}$.

Besides the regular graph case, we can also consider irregular code ensembles.
Let  $\lambda$ and $\rho$ denote the finite order {\it edge degree
distribution} polynomials
    \begin{eqnarray}
    \lambda(x)&=&\sum_k \lambda_kx^{k-1}\nonumber\\
    \rho(x)&=&\sum_k \rho_kx^{k-1},\nonumber
    \end{eqnarray}
where $\lambda_k$ or $\rho_k$ is the fraction of edges connecting to a degree
$k$ variable or check node, respectively. By assigning equal probability to
each possible configuration of irregular bipartite graphs with degree
distributions $\lambda$ and $\rho$ (similarly to the regular case), we obtain
the equiprobable, irregular, bipartite graph ensemble ${\mathcal C}^n(\lambda,
\rho)$. For example: ${\mathcal C}^n(3,6)={\mathcal C}^n(x^2,x^5)$.

%
%
\subsection{Message Passing Algorithms \& Belief Propagation}
The message passing decoding algorithm is a distributed algorithm such that
each variable/check node has a processor, which takes all incoming messages
from its neighbors as inputs, and outputs new messages back to all its
neighbors. The algorithm can be completely specified by the variable and check
node message maps, $\Psi_v$ and $\Psi_c$, which may or may not be stationary
(i.e., the maps remain the same as time evolves) or uniform (i.e.,
node-independent). The message passing algorithm can be executed sequentially
or in parallel depending on the order of the activations of different node
processors. Henceforth, we consider only parallel message passing algorithms
complying with the {\it extrinsic} principle (adapted from turbo codes), i.e.\
the new message sending to node $i$ (or $j$) does not depend on the received
message from the same node $i$ (or $j$) but depends only on other received
messages.

A belief propagation algorithm is a message passing algorithm whose  variable
and check node message maps are derived from Pearl's inference network
\cite{Pearl88}. Under the cycle-free assumption on the inference network,
belief propagation calculates the exact marginal {\it a posteriori}
probabilities, and thus we obtain the optimal maximum {\it a posteriori}
probability (MAP) decisions. Let $m_0$ denote the initial message from the
variable nodes, and $\{m_k\}$ denote the messages from its neighbors excluding
that from the destination node. The entire belief propagation algorithm with
messages representing the corresponding log likelihood ratio (LLR) is as
follows:{\small
    \begin{eqnarray}
    m_0&:=&\ln\frac{\PP(y_i|x_i=0)}{\PP(y_i|x_i=1)}\nonumber\\
    \Psi_v(m_0,m_1,\cdots,m_{d_v-1})&:=&\sum_{j=0}^{d_v-1}m_j\label{eq:Psi-v}\\
    \Psi_c(m_1,\cdots,m_{d_c-1})&:=&\ln\left(\frac{1+\prod_{i=1}^{d_c-1}\tanh\frac{m_i}{2}}{1-\prod_{i=1}^{d_c-1}\tanh\frac{m_i}{2}}\right).\label{eq:Psi-c}
    \end{eqnarray}}\cmpt
We note that the belief propagation algorithm is based only on the cycle-free
assumption\footnote{An implicit assumption will be revisited in
Section~\ref{subsec:side-results-BP}.} and is actually independent of the
channel model. The initial message $m_0$ depends only on the single-bit LLR
function and  can be calculated under non-symmetric $f(y_i|x_i)$. As a result,
the belief propagation algorithm remains the same for memoryless,
symbol-dependent channels.

\begin{itemize}
\item {\it Example:} For BASCs,
    \begin{eqnarray}
    m_0=\begin{cases}
    \ln\frac{1-\epsilon_0}{\epsilon_1} & \text{if $y_i=0$}\\
    \ln\frac{\epsilon_0}{1-\epsilon_1} & \text{if $y_i=1$}
 \end{cases}.\nonumber
    \end{eqnarray}
\end{itemize}
We assume that the belief propagation is  executed in parallel and each {\it
iteration} is a ``round" in which all variable nodes send messages to all check
nodes and then the check nodes send messages back. We use $l$ to denote the
number of iterations that have been executed.

\subsection{Density Evolution}
For a symmetric channel and any message-passing algorithm, the probability
density of the transmitted messages in each iteration can be calculated
iteratively with a concrete theoretical foundation \cite{RichardsonUrbanke01a}.
The iterative formula and related theorems are termed ``density evolution."
Since the belief propagation algorithm performs extremely well under most
circumstances and is of great importance, sometimes the term ``density
evolution" is reserved for the corresponding analytical method for belief
propagation algorithms.

%
%
\section{Density Evolution: New Iterative Formula\label{sec:iterative-formula}}
In what follows, we use the belief propagation algorithm as the
illustrative example for our new iterative density evolution
formula.

With the assumption of channel symmetry and the inherent symmetry of the parity
check equations in LDPC codes, the probability density of the messages in any
symmetric message passing algorithm will be codeword independent, i.e., for
different codewords, the densities of the messages passed differ only in
parities, but all of them are of the same shape [\Lemma{1},
\cite{RichardsonUrbanke01a}].

 In the symbol-dependent setting, symmetry of
the channel may not hold. Even though the belief propagation mappings remain
the same for asymmetric channels, the densities of the messages for different
transmitted codewords are of different shapes and the density for the all-zero
codeword cannot represent the behavior when other codewords are transmitted. To
circumvent this problem, we {\it average} the density of the messages over all
valid codewords. However, directly averaging over all codewords takes $2^{n-m}$
times more computations, which ruins the efficiency of the iterative formula
for density evolution. Henceforth, we provide a new iterative formula for the
codeword-averaged density evolution which increases the number of computations
only by a constant factor; the corresponding theoretical foundations are
provided in this section and in Section~\ref{sec:theorems}.

\subsection{Preliminaries}
We consider the density of the message passed from variable node $i$ to check
node $j$. The probability density of this message is denoted by
$P^{(l)}_{(i,j)}({\mathbf x})$ where the superscript $l$ denotes the {$l$-th}
iteration and the appended argument $\mathbf x$ denotes the actual transmitted
codeword. For example, $P^{(1)}_{(i,j)}({\mathbf 0})$ is the density of the
initial message $m_0$ from variable node~$i$ to check node~$j$ assuming the
all-zero codeword is transmitted. $P^{(2)}_{(i,j)}({\mathbf 0})$ is the density
from $i$ to $j$ in the second iteration, and so on.  We also denote by
$Q^{(l)}_{(j,i)}({\mathbf x})$ the density of the message from check node~$j$
to variable node~$i$ in the $l$-th iteration.

\begin{figure}[t]
\begin{center}
\begin{tabular}{cc}
\parbox[c]{3cm}
{\includegraphics[width=2cm, keepaspectratio=true]{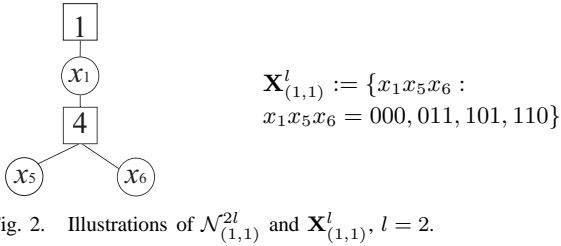}} &
\parbox[l]{5cm}{
{\small ${\mathbf X}^l_{(1,1)}:=\{x_1x_5x_6:$\\
~\hspace{.4cm}$x_1x_5x_6=000,011,101,110 \}$}}
\end{tabular}
\caption{Illustrations of  ${\mathcal N}^{2l}_{(1,1)}$ and ${\mathbf
X}^l_{(1,1)}$, $l=2$.} \label{fig:codeword-projection}
\end{center}
\end{figure}

With the assumption that the corresponding graph is tree-like until depth
$2(l-1)$, we define the following quantities.
\Figure{\ref{fig:codeword-projection}} illustrates these quantities for the
code in \Figure{\ref{fig:codebook}} with $i=j=1$ and $l=2$.
    \begin{itemize}
    \item ${\mathcal N}^{2l}_{(i,j)}$ denotes the
        tree-like subset of the  graph\footnote{The calligraphic $\mathcal V$ in $G=({\mathcal V},{\mathcal E})$ denotes the set of all vertices,
        including both variable nodes and check nodes. Namely, a node $v\in\mathcal V$ can be a variable/check node.}
        $G=({\mathcal V},{\mathcal E})$ with root edge $(i,j)$ and depth $2(l-1)$, named as the supporting tree.
         A formal definition is: ${\mathcal N}^{2l}_{(i,j)}$
        is the subgraph induced by ${\mathcal V}^{2l}_{(i,j)}$, where
        {\small
        \begin{eqnarray}
        {\mathcal V}^{2l}_{(i,j)}:=\{v\in{\mathcal
        V}:d(v,i)=d(v,j)-1\in[0,2(l-1)]\}, \label{eq:definition-of-tree}
        \end{eqnarray}}\cmpt
        where $d(v,i)$ is the shortest distance between node $v$ and variable node
        $i$. In other words, ${\mathcal N}^{2l}_{(i,j)}$ is the
        depth $2(l-1)$ tree spanned from edge $(i,j)$.
         Let $\left|{\mathcal
        N}^{2l}_{(i,j)}\right|_V$ denote the number of variable nodes in
        ${\mathcal N}^{2l}_{(i,j)}$ (including variable node~$i$).
        $\left|{\mathcal
        N}^{2l}_{(i,j)}\right|_C$ denotes the number of check nodes in
        ${\mathcal N}^{2l}_{(i,j)}$ (check node~$j$ is excluded by definition).
    \item ${\mathbf X}=\left\{{\mathbf x}\in\{0,1\}^n:{\mathbf A}{\mathbf
        x}={\mathbf 0}\right\}$ denotes the set of all valid
        codewords, and the information source selects each codeword equiprobably from $\mathbf
        X$.
    \item ${\mathbf x}|_i$ and
        ${\mathbf x}|_{{\mathcal N}^{2l}_{(i,j)}}$ are the projections of
        codeword ${\mathbf x}\in{\mathbf X}$ on bit $i$ and on the variable nodes in the supporting tree ${\mathcal
        N}^{2l}_{(i,j)}$, respectively.
    \item ${\mathbf X}^l_{(i,j)}$ denotes the set of all strings of length
        $\left|{\mathcal
        N}^{2l}_{(i,j)}\right|_V$ satisfying the $\left|{\mathcal
        N}^{2l}_{(i,j)}\right|_C$
        check node constraints in ${\mathcal N}^{2l}_{(i,j)}$.
        ${\mathbf x}^l$ denotes any element of ${\mathbf X}^l_{(i,j)}$ (the subscript $(i,j)$ is omitted if there is no ambiguity).
        The connection between ${\mathbf X}$, the valid codewords, and ${\mathbf
        X}^l_{(i,j)}$, the tree-satisfying strings, will be clear in the
        following remark and in
        \Definition{\ref{def:perfect-projection}}.
    \item For any set of codewords (or strings) ${\mathbf W}$, the average
        operator $\langle\cdot\rangle_{\mathbf W}$ is defined as:
            \begin{eqnarray}
            \left\langle g({\mathbf x})\right\rangle_{\mathbf W}=\frac{1}{|{\mathbf
            W}|}\sum_{{\mathbf x}\in{\mathbf W}} g({\mathbf
            x}).\nonumber
            \end{eqnarray}
    \item With a slight abuse of notation for $P^{(l)}_{(i,j)}(x)$, we define
            \begin{eqnarray}
            P_{(i,j)}^{(l)}(x)
            &:=&\left\langle P_{(i,j)}^{(l)}({\mathbf x})\right\rangle_{\{{\mathbf x}\in{\mathbf X}:{\mathbf
            x}|_{i}=x\}}\nonumber\\
            P_{(i,j)}^{(l)}({\mathbf x}^l)
            &:=&\left\langle P_{(i,j)}^{(l)}({\mathbf x})\right\rangle_{\{{\mathbf x}\in{\mathbf X}:{\mathbf x}|_{{\mathcal N}^{2l}_{(i,j)}}={\mathbf x}^l\}}.\nonumber
            \end{eqnarray}
            Namely, $P_{(i,j)}^{(l)}(x)$ and $P_{(i,j)}^{(l)}({\mathbf x}^l)$
            denote the density averaged over all compatible codewords with
            projections being $x$ and ${\mathbf x}^l$, respectively.
    \end{itemize}

        {\it Remark:} For any tree-satisfying string ${\mathbf x}^l\in {\mathbf X}^l_{(i,j)}$, there may or may not be a codeword $\mathbf x$ with
        projection ${\mathbf x}|_{{\mathcal N}^{2l}_{(i,j)}}={\mathbf
        x}^l$, since the codeword $\mathbf x$ must satisfy {\it all} check
        nodes, but the string ${\mathbf x}^l$  needs to satisfy only $\left|{\mathcal N}^{2l}_{(i,j)}\right|_C$ constraints. Those check nodes outside ${\mathcal
        N}^{2l}_{(i,j)}$ may limit the projected space ${\mathbf
X}|_{{\mathcal N}^{2l}_{(i,j)}}$ to
        a strict subset of ${\mathbf X}^l_{(i,j)}$. For example, the second row of ${\mathbf Ax}={\mathbf 0}$ in \Figure{\ref{fig:codebook}}
         implies $x_6=0$. Therefore
         two of the four elements of ${\mathbf X}^l_{(1,1)}$ in \Figure{\ref{fig:codeword-projection}} are
         invalid/impossible
         projections of $\mathbf x\in\mathbf X$ on ${\mathcal N}^{2l}_{(1,1)}$. Thus ${\mathbf
X}|_{{\mathcal N}^{2l}_{(1,1)}}$ is a proper subset of ${\mathbf
X}^l_{(1,1)}$.

To capture this phenomenon, we introduce the notion of a {\it perfectly
projected} ${\mathcal N}^{2l}_{(i,j)}$.
    \begin{definition}[Perfectly Projected ${\mathcal
    N}^{2l}_{(i,j)}$]\label{def:perfect-projection}
     The supporting tree
    ${\mathcal N}^{2l}_{(i,j)}$ is perfectly projected, if for any ${\mathbf  x}^l\in{\mathbf X}^l_{(i,j)}$,
        \begin{eqnarray}
        \frac{\left|\{{\mathbf x}\in {\mathbf X}: {\mathbf x}|_{{\mathcal N}^{2l}_{(i,j)}}={\mathbf
        x}^l\}\right|}
        {|{\mathbf X}|}=\frac{1}{\left|{\mathbf
        X}^l_{(i,j)}\right|}.\label{eq:def-perfect-proj}
        \end{eqnarray}
    That is, if we choose $\mathbf x\in\mathbf X$ equiprobably,
    ${\mathbf x}|_{{\mathcal N}^{2l}_{(i,j)}}$ will appear uniformly among
    all elements in ${\mathbf x}^l\in{\mathbf X}^l_{(i,j)}$. Thus by looking only at
    the projections  on
    ${\mathcal N}^{2l}_{(i,j)}$, it is as if we are choosing ${\mathbf x}^l$ from ${\mathbf X}^l_{(i,j)}$ equiprobably and there are only
    $\left|{\mathcal N}^{2l}_{(i,j)}\right|_C$ check node constraints and no
    others.
    \end{definition}
The example in \Figures{\ref{fig:codebook} {\rm
and}~\ref{fig:codeword-projection}} is obviously not perfectly projected.

Since the message emitted from node $i$ to $j$ in the $l$-th iteration depends
only on the received signals of the supporting tree, $\mathbf y|_{{\mathcal
N}^{2l}_{(i,j)}}$, the codeword-dependent $P_{(i,j)}^{(l)}({\mathbf x})$
actually depends only on the projection ${\mathbf x}|_{{\mathcal
N}^{2l}_{(i,j)}}$, not on the entire codeword $\mathbf x$. That is
    \begin{eqnarray}
    P_{(i,j)}^{(l)}({\mathbf x})=P_{(i,j)}^{(l)}({\mathbf x}|_{{\mathcal
N}^{2l}_{(i,j)}}).\label{eq:restriction1}
    \end{eqnarray}
An immediate implication of ${\mathcal N}^{2l}_{(i,j)}$ being a perfect
projection and (\ref{eq:restriction1}) is
    \begin{eqnarray}
    P_{(i,j)}^{(l)}(x)&:=&\left\langle P_{(i,j)}^{(l)}({\mathbf x})\right\rangle_{\{{\mathbf x}\in{\mathbf X}:{\mathbf
            x}|_{i}=x\}}\nonumber\\
            &=&\frac{1}{\left|\{{\mathbf x}\in{\mathbf X}:{\mathbf
            x}|_{i}=x\}\right|}\sum_{\{{\mathbf x}\in{\mathbf X}:{\mathbf
            x}|_{i}=x\}}P_{(i,j)}^{(l)}({\mathbf x})\nonumber\\
            &=&\frac{1}{|\{{\mathbf x}\in{\mathbf X}:{\mathbf
            x}|_{i}=x\}|}\nonumber\\
            &&~\cdot\left|\{{\mathbf x}\in{\mathbf X}:{\mathbf x}|_{{\mathcal N}^{2l}_{(i,j)}}={\mathbf
            x}^l, {\mathbf x}^l|_i=x\}\right|\nonumber\\
            &&~\cdot
            \sum_{\{{\mathbf x}^l\in{\mathbf X}^l_{(i,j)}, {\mathbf x}^l|_i=x \}} P_{(i,j)}^{(l)}({\mathbf x}^l)
            \nonumber\\
    &=&\left\langle P_{(i,j)}^{(l)}({\mathbf
    x}^l)\right\rangle_{\{{\mathbf x}^l\in {\mathbf X}^l_{(i,j)}: {\mathbf
    x}^l|_i=x\}}.\label{eq:restriction3}
    \end{eqnarray}
Because of these two useful properties, (\ref{eq:restriction1}) and
(\ref{eq:restriction3}), throughout this subsection we assume that ${\mathcal
N}^{2l}_{(i,j)}$ is perfectly projected. The convergence of ${\mathcal
N}^{2l}_{(i,j)}$ to a perfect projection in probability is dealt with in
Section~\ref{sec:theorems}. We will have all the preliminaries necessary for
deriving the new density evolution after introducing the following
self-explanatory lemma.
    \begin{lemma}[Linearity of Density Transformation]
    For any random variable $A$ with distribution $P_A$, if $g:A\mapsto
    g(A)$ is measurable, then $B=g(A)$ is a random variable with distribution $P_B=T_g(P_A):=P_A\circ
    g^{-1}$.
    Furthermore, the density transformation $T_g$ is linear. I.e.\ if $P_{B}=T_g(P_{A})$ and $Q_{B}=T_g(Q_{A})$, then $\alpha P_{B}+(1-\alpha)Q_{B}=T_g(\alpha
    P_{A}+(1-\alpha)Q_{A})$, $\forall \alpha\in[0,1]$.\label{lem:linearity-of-distribution}
    \end{lemma}

\subsection{New Formula}
In the $l$-th iteration, the probability of sending an incorrect message
(averaged over all possible codewords) from variable node $i_0$ to check node
$j_0$ is {\small
    \begin{eqnarray}
    p_e^{(l)}(i_0,j_0)&=&\frac{1}{|{\mathbf X}|}\left(\sum_{\{{\mathbf x}\in{\mathbf X}: {\mathbf
        x}|_{i_0}=0\}}\int_{m=-\infty}^0 P^{(l)}_{(i_0,j_0)}({\mathbf
        x})(dm)\right.\nonumber\\
        &&~~~~~~+\left.
        \sum_{\{{\mathbf x}\in{\mathbf X}: {\mathbf
            x}|_{i_0}=1\}}\int_{m=0}^\infty P^{(l)}_{(i_0,j_0)}({\mathbf
            x})(dm)\right)\nonumber\\
    &=&\frac{1}{2}\left(\int_{m=-\infty}^0 P^{(l)}_{(i_0,j_0)}(0)(dm)\right.\nonumber\\
    &&~~~~+\left.\int_{m=0}^\infty P^{(l)}_{(i_0,j_0)}(1)(dm)\right).\label{eq:error-probability}
    \end{eqnarray}}\cmpt
Motivated by  (\ref{eq:error-probability}), we concentrate on
finding an iterative formula for the density pair $
P^{(l)}_{(i_0,j_0)}(0)$ and $P^{(l)}_{(i_0,j_0)}(1)$. Throughout
this section, we also assume ${\mathcal N}^{2l}_{(i_0,j_0)}$ is
tree-like (cycle-free) and perfectly projected.

\begin{figure}[t]
\centering \includegraphics[width=8.5cm,
keepaspectratio=true]{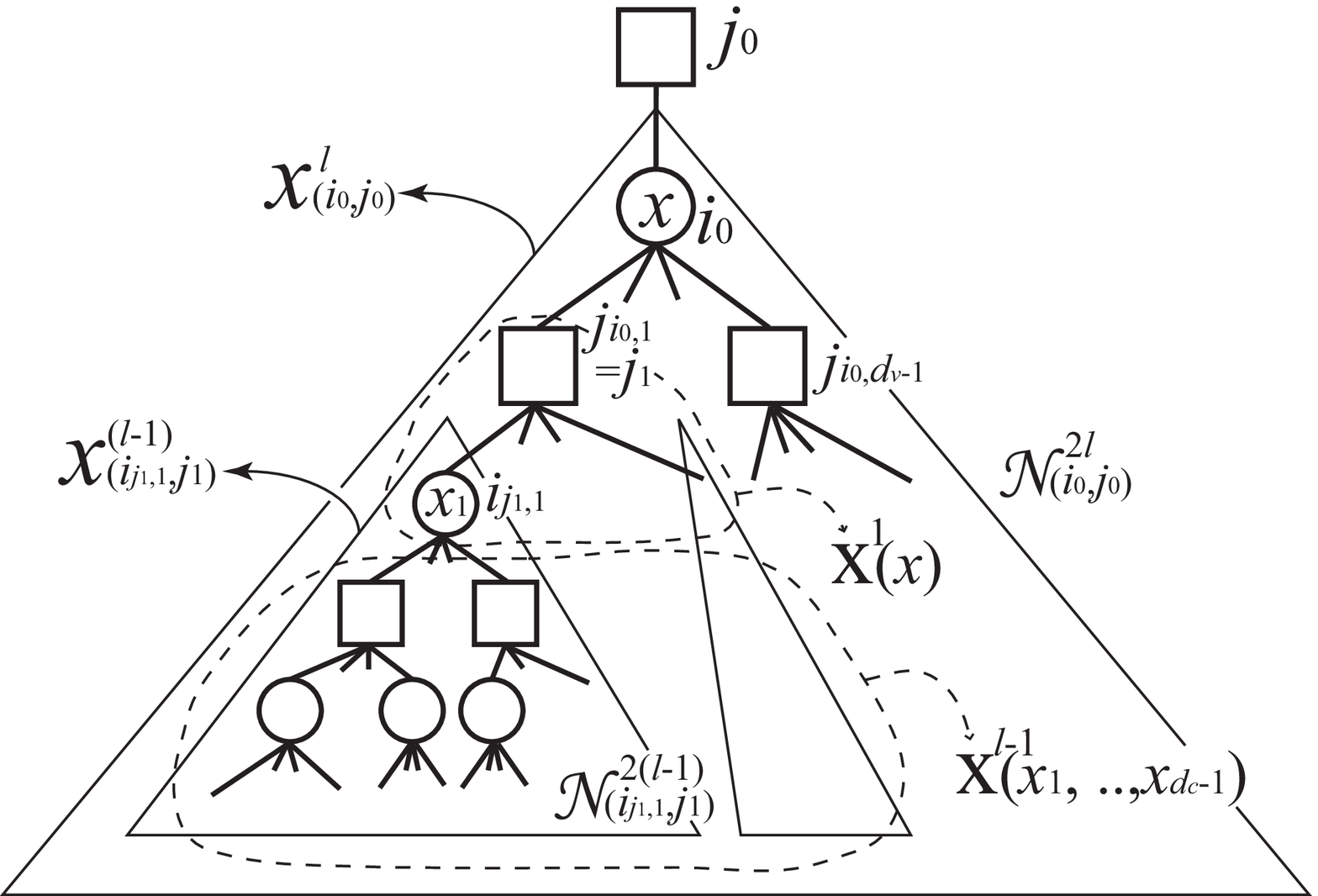}\caption{Illustrations of
various quantities used in Section~\ref{sec:iterative-formula}.}
\label{fig:Quantities-illustrations}
\end{figure}

Let $1_{\{\cdot\}}$ denote the indicator function. By an auxiliary function
$\gamma(m)$:
    \begin{eqnarray}
    \gamma(m)&:=&\left(1_{\{m\leq0\}},\ln\coth\left|\frac{m}{2}\right|\right),\label{eq:domain-gamma}
    \end{eqnarray}
and letting the domain of the first coordinate of $\gamma(m)$ be $\GF(2)$,
Eq.~(\ref{eq:Psi-c}) for $\Psi_c$ can be written as
    \begin{eqnarray}
    \Psi_c(m_1,\cdots,m_{d_c-1})&=&\gamma^{-1}\left(\sum_{v=1}^{d_c-1}\gamma(m_v)\right).\label{eq:definition-of-gamma}
    \end{eqnarray}
By (\ref{eq:Psi-v}), (\ref{eq:definition-of-gamma}), and the independence among
the input messages, the classical density evolution for belief propagation
algorithms (Eq.~(9) in \cite{RichardsonShokrollahiUrbanke01}) is as follows.
{\small
    \begin{eqnarray}
    P_{(i_0,j_0)}^{(l)}({\mathbf
        x})&=&P_{(i_0,j_0)}^{(0)}({\mathbf
        x})\otimes \left(\bigotimes_{c=1}^{d_v-1}Q_{(j_{i_0,c},i_0)}^{(l-1)}({\mathbf
        x})\right)\label{eq:density-evolution-P}\\
    Q_{(j_{i_0,c},i_0)}^{(l-1)}({\mathbf
        x})&=&\Gamma^{-1}\left(\bigotimes_{v=1}^{d_c-1}\Gamma\left( P_{(i_{j,v},j_{i_0,c})}^{(l-1)}({\mathbf
        x})\right)\right),\label{eq:density-evolution-Q}
    \end{eqnarray}}\cmpt
where $\otimes$ denotes the convolution operator on probability density
functions, which can be implemented efficiently  using the  Fourier transform.
$\Gamma:=T_\gamma$ is the density transformation functional based on $\gamma$,
defined in \Lemma{\ref{lem:linearity-of-distribution}}.
\Figure{\ref{fig:Quantities-illustrations}} illustrates many helpful quantities
used in (\ref{eq:density-evolution-P}), (\ref{eq:density-evolution-Q}), and
throughout this section.

By (\ref{eq:restriction1}), (\ref{eq:density-evolution-P}), and the perfect
projection assumption, we have
    \begin{eqnarray}
    P_{(i_0,j_0)}^{(l)}({\mathbf
    x}^l)=P_{(i_0,j_0)}^{(0)}({\mathbf
    x}|_{i_0})\otimes\left(\bigotimes_{c=1}^{d_v-1}Q_{(j_{i_0,c},i_0)}^{(l-1)}({\mathbf
    x}^l)\right).\label{eq:classic-result}
    \end{eqnarray}
Further simplification can be made such that
    \begin{eqnarray}
    &&\hspace{-.7cm}P_{(i_0,j_0)}^{(l)}(x)\nonumber\\
    &\stackrel{(a)}{=}&\left\langle P_{(i_0,j_0)}^{(l)}({\mathbf
        x}^l)\right\rangle_{\{{\mathbf
        x}^l:{\mathbf x}^l|_{i_0}=x\}}\nonumber\\
    &\stackrel{(b)}{=}&\left\langle P_{(i_0,j_0)}^{(0)}(x)\otimes\left(\bigotimes_{c=1}^{d_v-1}Q_{(j_{i_0,c},i_0)}^{(l-1)}({\mathbf
        x}^l)\right)\right\rangle_{\{{\mathbf
        x}^l:{\mathbf x}^l|_{i_0}=x\}}\nonumber\\
    &\stackrel{(c)}{=}&P_{(i_0,j_0)}^{(0)}(x)\otimes \left\langle\bigotimes_{c=1}^{d_v-1}Q_{(j_{i_0,c},i_0)}^{(l-1)}({\mathbf
        x}^l)\right\rangle_{\{{\mathbf
        x}^l:{\mathbf x}^l|_{i_0}=x\}}\nonumber\\
    &\stackrel{(d)}{=}&
        P_{(i_0,j_0)}^{(0)}(x)\otimes\left(\bigotimes_{c=1}^{d_v-1}\left\langle Q_{(j_{i_0,c},i_0)}^{(l-1)}({\mathbf
        x}^l)\right\rangle_{\{{\mathbf
        x}^l:{\mathbf x}^l|_{i_0}=x\}}\right)\nonumber\\
    &\stackrel{(e)}{=}&P_{(i_0,j_0)}^{(0)}(x)\otimes\left(\left\langle Q_{(j_{i_0,1},i_0)}^{(l-1)}({\mathbf
        x}^l)\right\rangle_{\{{\mathbf
        x}^l:{\mathbf x}^l|_{i_0}=x\}}\right)^{\otimes(d_v-1)},\nonumber\\ \label{eq:iterative-formula-1}
    \end{eqnarray}
where (a) follows from (\ref{eq:restriction3}), (b) follows from
(\ref{eq:classic-result}), and (c) follows from the linearity of convolutions.
The fact that the sub-trees generated by edges $(j_{i_0,c},i_0)$ are completely
disjoint implies that, by the perfect projection assumption on ${\mathcal
N}^{2l}_{(i_0,j_0)}$, the distributions of strings on different sub-trees are
independent. As a result, the average of the convolutional products (over these
strings) equals the convolution of the averaged distributions, yielding (d).
Finally (e) follows from the fact that the distributions of messages from
different subtrees are identical according to the perfect projection
assumption.

To simplify $\left\langle Q_{(j_{i_0,1},i_0)}^{(l-1)}({\mathbf
        x}^l)\right\rangle_{\{{\mathbf
        x}^l:{\mathbf x}^l|_{i_0}=x\}}$, we need to define some new notation. We use $j_1$ to represent $j_{i_0,1}$ for simplicity.
        Denote by
$\left\{{\mathcal N}^{2(l-1)}_{(i_{j_{1},v},j_{1})}\right\}_{v\in[1,d_c-1]}$
the collection of all $d_c-1$ subtrees rooted at $(i_{j_1,v},j_{1})$,
$v\in[1,d_c-1]$, and by ${\mathbf X}^{l-1}_{(i_{j_1,v}, j_{1})}$ the strings
compatible to ${\mathcal N}^{2(l-1)}_{(i_{j_1,v},j_{1})}$. We can then consider
    \begin{eqnarray}
        {\mathbf
            X}^1(x)=\left\{(x_1,\cdots,x_{d_c-1}):\left(\sum_{v=1}^{d_c-1}x_v\right)+x=0\right\}\nonumber
    \end{eqnarray}
    containing the strings satisfying parity check constraint $j_1$ given
    $x_{i_0}=x$, and
    \begin{eqnarray}
        &&\hspace{-.7cm}{\mathbf X}^{l-1}(x_1,\cdots,x_{d_c-1})\nonumber\\
        &:=&\left\{({\mathbf
                x}^{l-1}_{(i_{j_1,1},j_1)},\cdots,{\mathbf x}^{l-1}_{(i_{j_1,d_c-1},j_1)}):\right.\nonumber\\
        &&~~~~~~~~\left.{\mathbf
                x}^{l-1}_{(i_{j_1,1},j_1)}|_{i_{j_1,1}}=x_1,\cdots,\right.\nonumber\\
                &&~~~~~~~~\left.{\mathbf
                x}^{l-1}_{(i_{j_1,d_c-1},j_1)}|_{i_{j_1,d_c-1}}=x_{d_c-1}\right\}\nonumber
    \end{eqnarray}
is the collection of the concatenations of substrings, in which the leading
symbols of the substrings are
   $(x_1,\cdots,x_{d_c-1})$. All these quantities are illustrated in \Figure{\ref{fig:Quantities-illustrations}}.

Note the following two properties: (i) For any $v$, the message $m_v$ from
variable $i_{j_1,v}$ to check node $j_1$ depends only on ${\mathbf
                x}^{l-1}_{(i_{j_1,v},j_1)}$;
                 and (ii) With the
leading symbols
                $\{x_v\}_{v\in[1,d_c-1]}$ fixed and the perfect projection assumption, the projection on the strings $\left\{{\mathbf
                x}^{l-1}_{(i_{j_1,v},j_1)}\right\}_{v\in[1,d_c-1]}$ are independent, and thus the averaged convolution of densities is equal to the
     convolution of the averaged densities.
By repeatedly applying \Lemma{\ref{lem:linearity-of-distribution}} and the
above two properties, we have {\small
\begin{eqnarray}
        &&\left\langle Q_{(j_{i_0,1},i_0)}^{(l-1)}({\mathbf
            x}^l)\right\rangle_{\{{\mathbf
            x}^l:{\mathbf x}^l|_{i_0}=x\}}\nonumber\\
        && = \left\langle           \Gamma^{-1}\left(\bigotimes_{v=1}^{d_c-1}\Gamma\left( P_{(i_{j,v},j_{i_0,c})}^{(l-1)}({\mathbf
        x}^l)\right)\right)\right\rangle_{\{{\mathbf
            x}^l:{\mathbf x}^l|_{i_0}=x\}}\nonumber\\
        && = \left\langle           \Gamma^{-1}\left(\bigotimes_{v=1}^{d_c-1}\Gamma\left( P_{(i_{j,v},j_{i_0,c})}^{(l-1)}({\mathbf
                x}^{l-1}_{(i_{j_1,v},j_1)})\right)\right)\right\rangle_{\{{\mathbf
            x}^l:{\mathbf x}^l|_{i_0}=x\}}\nonumber\\
        &&=\Gamma^{-1}\left(\frac{1}{2^{d_c-2}}\sum_{{\mathbf x}^1\in{\mathbf
            X}^1(x)}\right.\nonumber\\
        &&\hspace{1.5cm}\left.    \left\langle \bigotimes_{v=1}^{d_c-1}\Gamma\left( P_{(i_{j,v},j_{i_0,c})}^{(l-1)}({\mathbf
                x}^{l-1}_{(i_{j_1,v},j_1)})\right)\right\rangle_{{\mathbf X}^{l-1}({\mathbf
        x}^1)}\right)\nonumber\\
        &&=\Gamma^{-1}\left(\frac{1}{2^{d_c-2}}\sum_{{\mathbf x}^1\in{\mathbf
            X}^1(x)}    \bigotimes_{v=1}^{d_c-1}\right.\nonumber\\
            &&\hspace{2cm}\left.\Gamma\left( \left\langle P_{(i_{j,v},j_{i_0,c})}^{(l-1)}({\mathbf
                x}^{l-1}_{(i_{j_1,v},j_1)})\right\rangle_{{\mathbf X}^{l-1}({\mathbf
        x}^1)}\right)\right)\nonumber\\
        &&=\Gamma^{-1}\left(\frac{1}{2^{d_c-2}}\sum_{{\mathbf x}^1\in{\mathbf
            X}^1(x)}    \bigotimes_{v=1}^{d_c-1}\Gamma\left(P_{(i_{j,v},j_{i_0,c})}^{(l-1)}(x_v)\right)\right)\label{eq:chk-iteration-formula}
\end{eqnarray}}

By (\ref{eq:iterative-formula-1}), (\ref{eq:chk-iteration-formula}), and
dropping the subscripts during the density evolution, a new density evolution
formula for $P^{(l)}(x)$, $\forall x=0,1$, is as follows.{\small
\begin{eqnarray}
    P^{(l)}(x) &=&P^{(0)}(x)\otimes
        \left(Q^{(l-1)}(x)\right)^{\otimes(d_v-1)}\nonumber\\
    Q^{(l-1)}(x)&=&
            \Gamma^{-1}\left(\frac{1}{2^{d_c-2}}\sum_{{\mathbf x}^1\in{\mathbf X}^1(x)}
            \bigotimes_{v=1}^{d_c-1}
        \Gamma\left(P^{(l-1)}(x_v)\right)\right).
            \nonumber
\end{eqnarray}}
With the help of the linearity of distribution transformations and
convolutions, the above can be further simplified  and the desired efficient
iterative formulae become: {\footnotesize
    \begin{eqnarray}
    P^{(l)}(x) &=&P^{(0)}(x)\otimes
        \left(Q^{(l-1)}(x)\right)^{\otimes(d_v-1)}\nonumber\\
    Q^{(l-1)}(x)&=& \Gamma^{-1}\left(\left(\Gamma\left(\frac{P^{(l-1)}(0)+P^{(l-1)}(1)}{2}\right)\right)^{\otimes
        (d_c-1)}\right.\nonumber\\
         &&\left.  +(-1)^x\left(\Gamma\left(\frac{P^{(l-1)}(0)-P^{(l-1)}(1)}{2}\right)\right)^{\otimes (d_c-1)}
        \right).
\nonumber
    \end{eqnarray}}

The above formula can be easily generalized to the irregular code ensembles
${\mathcal C}^n(\lambda,\rho)$:{\small
    \begin{eqnarray}
    P^{(l)}(x) &=&P^{(0)}(x)\otimes
        \lambda\left(Q^{(l-1)}(x)\right)\nonumber\\
    Q^{(l-1)}(x)&=&
         \Gamma^{-1}\left(\rho\left(\Gamma\left(\frac{P^{(l-1)}(0)+P^{(l-1)}(1)}{2}\right)\right)\right.\nonumber\\
     &&\left.       +(-1)^x\rho\left(\Gamma\left(\frac{P^{(l-1)}(0)-P^{(l-1)}(1)}{2}\right)\right)
        \right),\nonumber\\
\label{eq:main-result}
    \end{eqnarray}}\cmpt
which has the same complexity as the classical density evolution for symmetric
channels.

{\it Remark:} The above derivation relies heavily on the perfect projection
assumption, which guarantees that uniformly averaging over all codewords is
equivalent to uniformly averaging over the tree-satisfying strings. Since the
tree-satisfying strings are well-structured and symmetric, we are on solid
ground to move the average inside the classical density evolution formula.
%
%
\section{Density Evolution: Fundamental Theorems\label{sec:theorems}}
As stated in Section~\ref{sec:iterative-formula}, the tree-like until depth
$2l$
 and the prefect projection assumptions are critical in our analysis. The use
 of codeword ensembles rather than fixed codes facilitates the
analysis but its relationship to fixed codes needs to be explored.  We restate
two necessary theorems from \cite{RichardsonUrbanke01a}, and give a novel
perfect projection convergence theorem, which is essential to our new density
evolution method. With these theorems, a concrete theoretical foundation will
be established.

\begin{theorem}[Convergence to the Cycle-Free Case,
\cite{RichardsonUrbanke01a}] Fix $l$, $i_0$, and $j_0$. For any
$(d_v,d_c)$, there exists a constant $\alpha>0$, such that for all
$n\in\NN$, the code ensemble ${\mathcal C}^n(d_v,d_c)$ satisfies
    \begin{eqnarray}
    \PP\left(\mbox{${\mathcal
    N}^{2l}_{(i_0,j_0)}$ is
    cycle-free}\right)\geq1-\alpha\left( \frac{\{(d_v-1)(d_c-1)\}^{2l}}{n}\right),\nonumber
    \end{eqnarray}\label{thm:cyle-free}
where  ${\mathcal N}^{2l}_{(i_0,j_0)}$ is the support tree as
defined by (\ref{eq:definition-of-tree}).
\end{theorem}
\begin{theorem}[Convergence to Perfect Projection in Prob.]
Fix $l, i_0$, and $j_0$. For any regular, bipartite, equiprobable
graph ensemble ${\mathcal C}^n(d_v,d_c)$, we have
\begin{eqnarray}
\PP\left(\mbox{${\mathcal
    N}^{2l}_{(i_0,j_0)}$ is perfectly projected}\right)=1-\bigorder(n^{-0.1}).\nonumber
\end{eqnarray}\label{thm:degree-of-freedom}
\end{theorem}

{\it Remark:} The above two theorems focus only on the properties
of equiprobable regular bipartite graph ensembles, and are
independent of the channel type of interest.

\begin{theorem}[Concentration to the Expectation,
\cite{RichardsonUrbanke01a}] With fixed transmitted codeword $\mathbf x$, let
$Z$ denote the number of wrong messages (those $m$'s such that $m(-1)^x<0$).
There exists a constant $\beta>0$ such that for any $\epsilon>0$, over the code
ensemble ${\mathcal C}^n(d_v,d_c)$ and the channel realizations $\mathbf y$, we
have
    \begin{eqnarray}
    \PP\left(\left|\frac{Z-\EE\{Z\}}{nd_v}\right|>\frac{\epsilon}{2}\right)\leq
    2e^{-\beta\epsilon^2n}.\label{eq:concentration-theorem}
    \end{eqnarray}
Furthermore, $\beta$ is independent of $f_{{\mathbf y}|{\mathbf x}}({\mathbf
y}|{\mathbf x})$, and thus is independent of $\mathbf
x$.\label{thm:inviditual-code-convergence}
\end{theorem}
{\it Theorem~\ref{thm:inviditual-code-convergence}} can easily be generalized
to
 symbol-dependent channels in the following corollary.
\begin{corollary}
Over the equiprobable codebook $\mathbf X$, the code ensemble\footnote{The only
valid codeword for {\it all} code instances of the ensemble is the all-zero
codeword. Therefore, a fixed bit string is in general not a valid codeword for
most instances of the code ensemble, which hampers the averaging over the code
ensemble. This, however, can be circumvented by the following construction. We
first use Gaussian elimination to index the codewords, $1,\cdots, 2^{nR}$, for
any code instance in the code ensemble. And we then fix the index instead of
the codeword. The statements and the proof of
\Theorem{\ref{thm:inviditual-code-convergence}} hold verbatim after this slight
modification.} ${\mathcal C}^n(d_v,d_c)$, and channel realizations $\mathbf y$,
(\ref{eq:concentration-theorem}) still holds.\label{cor:concentration}
\end{corollary}
\begin{proof} Since the constant $\beta$ in
\Theorem{\ref{thm:inviditual-code-convergence}}
is independent of the transmitted codeword $\mathbf x$, after averaging over
the equiprobable codebook $\mathbf X$, the inequality still holds. That is,
    \begin{eqnarray}
    \PP\left(\left|\frac{Z-\EE\{Z\}}{nd_v}\right|>\frac{\epsilon}{2}\right)
    =\EE_{\mathbf x}\left\{\PP\left(\left|\frac{Z-\EE\{Z\}}{nd_v}\right|>\left.\frac{\epsilon}{2}\right|{\mathbf
    x}\right)\right\}\nonumber\\
    \leq\EE_{\mathbf x}\left\{ 2e^{-\beta\epsilon^2n}\right\}=
    2e^{-\beta\epsilon^2n}.\nonumber
    \end{eqnarray}
\end{proof}

Now we have all the prerequisite of proving  the theoretical foundation of our
codeword-averaged density evolution.
\begin{theorem}[Validity of Codeword-Averaged DE]
Consider any regular, bipartite, equiprobable graph ensemble
${\mathcal C}^n(d_v,d_c)$ with  fixed $l$, $i_0$, and $j_0$.
$p_e^{(l)}(i_0,j_0)$ is derived from (\ref{eq:error-probability})
and the codeword-averaged density evolution after $l$ iterations.
The probability over equiprobable codebook $\mathbf X$, the code
ensemble ${\mathcal C}^n(d_v,d_c)$, and the channel realizations
$\mathbf y$, satisfies
\begin{eqnarray}
    \PP\left(\left|\frac{Z}{nd_v}-p_e^{(l)}(i_0,j_0)\right|>\epsilon
    \right)=e^{-\epsilon^2\bigorder(n)}, \forall \epsilon>0.\nonumber
    \end{eqnarray}
\end{theorem}

\vspace{.3cm}
\begin{proof} We note that
    $\frac{Z}{nd_v}$ is bounded between 0 and 1.
    By  observing that{\small
    \begin{eqnarray}
        &&\left(\frac{Z}{nd_v}\right)1\{\mbox{${\mathcal
    N}^{2l}_{(i_0,j_0)}$ is cycle-free and perfectly projected}\}\nonumber\\
    &&\leq \left( \frac{Z}{nd_v}\right)\nonumber\\
    &&\leq\left(\frac{Z}{nd_v}-1\right)1\{\mbox{${\mathcal
    N}^{2l}_{(i_0,j_0)}$ is cycle-free and perfectly projected}\}\nonumber\\
    &&~~~~~+1,\nonumber
    \end{eqnarray}}\cmpt
and using \Theorems{\ref{thm:cyle-free} {\rm and}~\ref{thm:degree-of-freedom}},
we have
$\lim_{n\rightarrow\infty}{\EE}\left\{\frac{Z}{nd_v}\right\}=p_e^{(l)}(i_0,j_0)$.
Then by \Corollary{\ref{cor:concentration}}, the proof is complete.
\end{proof}
The proof of \Theorem{\ref{thm:degree-of-freedom}} will be included in
\Append{\ref{app:proof-convergence}}

%
%
\section{Monotonicity, Symmetry, \& Stability\label{sec:related-theorems}}
In this section, we prove the monotonicity, symmetry, and stability of our
codeword-averaged density evolution method on belief propagation algorithms.
Since the codeword-averaged density evolution reduces to the traditional one
when the channel of interest is symmetric, the following theorems also reduce
to those (in \cite{RichardsonShokrollahiUrbanke01}
and~\cite{RichardsonUrbanke01a}) for symmetric channels.

%
%
\subsection{Monotonicity}
\begin{proposition}[Monotonicity with Respect to $l$] Let $p_e^{(l)}$ denote the bit error probability of the
codeword-averaged density evolution defined in
(\ref{eq:error-probability}). Then $p_e^{(l+1)}\leq p_e^{(l)}$,
for all $l\in\NN$.
\end{proposition}
\begin{proof}
We first note that the codeword-averaged approach can be viewed as
concatenating a bit-to-sequence random mapper with the observation channels,
and  the larger the tree-structure is, the more observation/information the
decision maker has. Since the BP decoder is the optimal MAP decoder for the
tree structure of interest, the larger the tree is, the smaller the error
probability will be. The proof is thus complete.
\end{proof}

\begin{proposition}[Monotonicity w.r.t. Degraded Channels] Let $f(y|x)$ and $g(y|x)$ denote two different channel models,
such that $g(y|x)$ is degraded with respect to (w.r.t.) $f(y|x)$. The
corresponding decoding error probabilities, $p_{e,f}^{(l)}$ and
$p_{e,g}^{(l)}$, are defined in (\ref{eq:error-probability}). Then for any
fixed $l$, we have $p_{e,f}^{(l)}\leq p_{e,g}^{(l)}$.
\end{proposition}
\begin{proof}
By taking the same point of view that the codeword-averaged
approach is a concatenation of  a bit-to-sequence random mapper
with the observation channels, this theorem can be easily proved
by the channel degradation argument.
\end{proof}
%

%
%
\subsection{Symmetry}
We will now show that even though the evolved density is derived from
non-symmetric channels, there are still some symmetry properties inherent in
the symmetric structure of belief propagation algorithms. We define the
symmetric distribution pair as follows.
\begin{definition}[Symmetric Distribution Pairs]
    Two probability measures ${\PP}$ and ${\mathsf Q}$ are a symmetric pair if for any integrable
    function $h$, we have
    \begin{eqnarray}
    \int h(m)d{\PP}(m)&=\int{ e^{-m}h(-m)d{\mathsf
    Q}(m)}.\nonumber
    \end{eqnarray}
    A distribution  ${\PP}_s$ is {\it self-symmetric} if $({\PP}_s, {\PP}_s)$
    is a symmetric pair.
\end{definition}
\begin{proposition}\label{thm:symmetric-density}
    Let $I(m):=-m$ be a parity reversing function, and let  $P^{(l)}(0)$ and  $P^{(l)}(1)$ denote the
    resulting density functions   from the codeword-averaged density evolution.
    Then $P^{(l)}(0)$ and $P^{(l)}(1)\circ I^{-1}$ are a
    symmetric pair for all $l\in\NN$.
\end{proposition}
{\it Remark:} In the symmetric channel case,  $P^{(l)}(0)$ and  $P^{(l)}(1)$
differ only in parity (\Lemma{1}, \cite{RichardsonUrbanke01a}). Thus,
$P^{(l)}(0)=P^{(l)}(1)\circ I^{-1}$ is self-symmetric [\Theorem{3} in
\cite{RichardsonShokrollahiUrbanke01}].

\begin{thmproof}{Proof:}
We note that by the equiprobable codeword distribution and the perfect
projection assumption, $P^{(l)}(0)$ and $P^{(l)}(1)$ act on the random variable
$m$, given by
    \begin{eqnarray}
        m:=\ln\frac{{\PP}(x=0|{\mathbf y}^l)}{{\PP}(x=1|{\mathbf y}^l)}=\ln\frac{{\PP}({\mathbf y}^l|x=0)}{{\PP}({\mathbf y}^l|x=1)},\nonumber
    \end{eqnarray}
where ${\mathbf y}^l$ is the received signal on the subset ${\mathcal N}^{2l}$
and ${\PP}$ is the distribution over channel realizations and equiprobable
codewords. Then by a change of  measure,
    \begin{eqnarray}
        &&\hspace{-.7cm}\int h(m) P^{(l)}(0)(dm)\nonumber\\&=&{\EE}_{x=0}\left\{h\left(\ln\frac{{\PP}({\mathbf y}^l|x=0)}{{\PP}({\mathbf y}^l|x=1)}\right)\right\}\nonumber\\
        &=&{\EE}_{x=1}\left\{\frac{{\PP}({\mathbf y}^l|x=0)}{{\PP}({\mathbf y}^l|x=1)}h\left(\ln\frac{{\PP}({\mathbf y}^l|x=0)}{{\PP}({\mathbf y}^l|x=1)}\right)\right\}\nonumber\\
        &=&\int e^m h(m) P^{(l)}(1)(dm).\label{eq:self-symm}
    \end{eqnarray}
This completes the proof.
\end{thmproof}
\begin{corollary}
    \begin{eqnarray}
    \langle P^{(l)}\rangle:=\frac{P^{(l)}(0)+P^{(l)}(1)\circ I^{-1}}{2}\nonumber
    \end{eqnarray}
is self-symmetric for all $l$, i.e. $(\langle P^{(l)}\rangle,\langle P^{(l)}\rangle)$ is a
symmetric pair.
\end{corollary}

%
%
\subsection{Stability}
Rather than  looking only at the error probability $p_e^{(l)}$ of the evolved
densities $P^{(l)}(0)$ and $P^{(l)}(1)$, we also focus on its Chernoff bound,
\begin{eqnarray}
CBP^{(l)}(x):=\int e^{-\frac{(-1)^xm}{2}}P^{(l)}(x)(dm).\nonumber
\end{eqnarray}
By letting $h(m)=e^{-\frac{m}{2}}$ and by (\ref{eq:self-symm}), we have
$CBP^{(l)}(0)=CBP^{(l)}(1)$. The averaged $\langle CBP^{(l)}\rangle$ then
becomes
\begin{eqnarray}
\langle CBP^{(l)}\rangle
&:=&\frac{1}{2}\left(CBP^{(l)}(0)+CBP^{(l)}(1)\right)\nonumber\\
&=&CBP^{(l)}(0)=CBP^{(l)}(1)\nonumber\\
&=&\int e^{\frac{-m}{2}}\langle P^{(l)}\rangle(dm).\label{eq:averaged-CBP}
\end{eqnarray}
We state three properties which can  easily be derived from the self-symmetry
of $\langle P^{(l)}\rangle$. Proofs can be found in
\cite{Khandekar02,RichardsonShokrollahiUrbanke01},
and~\cite{WangKulkarniPoor04}.
\begin{itemize}
\item $\langle CBP^{(l)}\rangle =\min_s\int e^{-s\cdot m}\langle
P^{(l)}\rangle(dm)$.
\item The density of $e^{-m/2}\langle
P^{(l)}\rangle(dm)$ is symmetric with respect to $m=0$.
\item $2p_e^{(l)}\leq \langle CBP^{(l)}\rangle \leq 2\sqrt{p_e^{(l)}(1-p_e^{(l)})}$. This justifies the use of $\langle CBP^{(l)}\rangle $ as
our performance measure.
\end{itemize}

Thus, we  consider $\langle CBP^{(l)}\rangle $, the Chernoff bound of
$p_e^{(l)}$. With the regularity assumption that $\int_{{\mathbf
R}}e^{sm}\langle P^{(0)}\rangle(dm)<\infty$ for all $s$ in some neighborhood of
zero, we state the necessary and sufficient stability conditions as follows.
\begin{theorem}[Sufficient Stability Condition]\label{thm:stability-suff}
Let $r:=\langle CBP^{(0)}\rangle=\int_{\mathbf R} e^{-m/2}\langle
P^{(0)}\rangle (dm)$. Suppose $\lambda_2\rho'(1)r<1$, and let $\epsilon^*$ be
the smallest strictly positive root of the following equation.
\begin{eqnarray}
\lambda(\rho'(1)\epsilon)r=\epsilon.\nonumber
\end{eqnarray}
If for some $l_0$, $\langle CBP^{(l_0)}\rangle<\epsilon^*$, then
 \begin{eqnarray}
\langle CBP^{(l)}\rangle=\begin{cases}
     {\bigorder}\left(\left(\lambda_2\rho'(1)r\right)^l\right) & \text{if $\lambda_2>0$}\\
    {\bigorder}\left(e^{-\bigorder\left((k_\lambda-1)^l\right)}\right) & \text{if $\lambda_2=0$}
 \end{cases},\nonumber
 \end{eqnarray}
where $k_\lambda=\min\{k:\lambda_k>0\}$. In both cases: $\lambda_2=0$ and
$\lambda_2>0$, $\lim_{l\rightarrow\infty}\langle CBP^{(l)}\rangle=0$.
\end{theorem}
\begin{corollary}
    For any noise distribution $f(y|x)$ with Bhattacharyya noise parameter $r:=\langle CBP^{(0)}\rangle$, if there is no $\epsilon\in(0,r)$ such that
    \begin{eqnarray}
        \lambda(\rho'(1)\epsilon)r=\epsilon,\nonumber
    \end{eqnarray}
    then ${\mathcal C}(\lambda,\rho)$ will have arbitrarily small bit
    error rate as $n$ tends to infinity. The corresponding $r$ can
    serve as an inner bound of the achievable region for general non-symmetric memoryless channels. Further discussion of
    finite dimensional bounds on the achievable region can be found in
    \cite{WangKulkarniPoor04}.
\end{corollary}
\begin{theorem}[Necessary Stability Condition]\label{thm:stability-nec}
Let $r:=\langle CBP^{(0)}\rangle$.
 If $\lambda_2\rho'(1)r>1$, then  $\lim_{l\rightarrow \infty}p^{(l)}_e>0$.
\end{theorem}
\begin{itemize}
\item {\it Remark 1:} $\langle CBP^{(0)}\rangle$ is the Bhattacharyya noise parameter and is related to the cutoff
rate $R_0$ by $R_0=1-\log_2(1+\langle CBP^{(0)}\rangle)$. Further discussion of
$\langle CBP^{(0)}\rangle$ for turbo-like and LDPC codes can be found in
\cite{JinMcEliece02,Khandekar02,WangKulkarniPoor04}.
\item {\it Remark 2:} The stability results are first stated in \cite{RichardsonShokrollahiUrbanke01}
without the convergence rate statement and the stability region $\epsilon^*$.
Since we focus on general asymmetric channels (with symmetric channels as a
special case), our convergence rate and stability region $\epsilon^*$ results
also apply to the symmetric channel case. Benefitting from considering its
Chernoff version, we will provide a simple proof, which did not appear in
\cite{RichardsonShokrollahiUrbanke01}.
\item {\it Remark 3:} $\epsilon^*$ can be used as a stopping criterion for the iterations of the density evolution.
Moreover, $\epsilon^*$ is lower bounded by
$\frac{1-\lambda_2\rho'(1)r}{\lambda(\rho'(1))r-\lambda_2\rho'(1)r}$, which is
a computationally efficient substitute for $\epsilon^*$.
\end{itemize}
\begin{thmproof}{Proof of \Theorem{\ref{thm:stability-suff}}:}
We define the Chernoff bound of the density of the messages emitting from check
nodes, $CBQ^{(l)}(x)$, in a fashion similar to $CBP^{(l)}(x)$:
\begin{eqnarray}
CBQ^{(l)}(x):=\int e^{-\frac{(-1)^xm}{2}}Q^{(l)}(x)(dm).\nonumber
\end{eqnarray}
First consider the case in which $d_c=3$. We then have
\begin{eqnarray}
    \Psi_{c}(m_1,m_2)&=&\ln\left(\frac{1+\tanh\frac{m_1}{2}\tanh\frac{m_2}{2}}{1-\tanh\frac{m_1}{2}\tanh\frac{m_2}{2}}\right)\nonumber\\
    &=&\ln\frac{e^{m_1}e^{m_2}+1}{e^{m_1}+e^{m_2}}.\nonumber
\end{eqnarray}
To simplify the analysis, we assume the all-zero codeword is transmitted and
then generalize the results to non-zero codewords. Suppose the distributions of
$m_1$ and $m_2$ are  $P^{(l)}_1(0)$ and $P^{(l)}_2(0)$, respectively. The
$CBQ^{(l)}(0)$ becomes
\begin{eqnarray}
&&\hspace{-.7cm}CBQ^{(l)}(0)\nonumber\\
&=&\int e^{-\frac{\Psi_{c}(m_1,m_2)}{2}}P^{(l)}_1(0)(dm_1)\times P^{(l)}_2(0)(dm_2)\nonumber\\
&=&\int \sqrt{\frac{e^{m_1}+e^{m_2}}{e^{m_1}e^{m_2}+1}}P^{(l)}_1(0)(dm_1)\times P^{(l)}_2(0)(dm_2)\nonumber\\
&\leq&\int \sqrt{e^{m_1}+e^{m_2}}P^{(l)}_1(0)(dm_1)\times P^{(l)}_2(0)(dm_2)\nonumber\\
&{\leq}&\int \sqrt{e^{m_1}}+\sqrt{e^{m_2}}P^{(l)}_1(0)(dm_1)\times P^{(l)}_2(0)(dm_2)\nonumber\\
&=&CBP^{(l)}_1(0)+CBP^{(l)}_2(0),\label{eq:density-inequality-of-Psi-c}
\end{eqnarray}
where the last inequality follows from the fact that $\forall
\alpha,\beta\geq0, \sqrt{\alpha+\beta}\leq\sqrt{\alpha}+\sqrt{\beta}$. Since
any check node with $d_c>3$ can be viewed as the concatenation of many check
nodes with $d_c=3$, by induction and by assuming the all-zero codeword is
transmitted, we have
\begin{eqnarray}
CBQ^{(l)}(0)\leq (d_c-1)CBP^{(l)}(0).\label{eq:CBQ-prelim-1}
\end{eqnarray}
Since $CBP^{(l)}(0)=CBP^{(l)}(1)$ as in (\ref{eq:averaged-CBP}), the averaging
over all possible codewords does not change (\ref{eq:CBQ-prelim-1}). By further
incorporating the check node degree polynomial $\rho$, we have
\begin{eqnarray}
\forall x\in\{0,1\}, ~CBQ^{(l)}(x)&\leq&\sum_{k}\rho_k(k-1)\left\langle
CBP^{(l)}\right\rangle\nonumber\\
&=&\rho'(1)\langle CBP^{(l)}\rangle.\nonumber
\end{eqnarray}
By (\ref{eq:main-result}) and the fact that the moment generating function of the convolution
equals the product of individual moment generating functions, we have
\begin{eqnarray}
CBP^{(l+1)}(x)&=&CBP^{(0)}(x)\sum_{k}\lambda_k\left(CBQ^{(l)}(x)\right)^{k-1}\nonumber\\
&\leq&CBP^{(0)}(x)\lambda\left(\rho'(1)\langle CBP^{(l)}\rangle\right),\nonumber
\end{eqnarray}
which is equivalent to
\begin{eqnarray}
\langle CBP^{(l+1)}\rangle&\leq&\langle CBP^{(0)}\rangle\lambda\left(\rho'(1)\langle
CBP^{(l)}\rangle\right).\label{eq:asymptotic-iteration-formula}
\end{eqnarray}
The sufficient stability theorem follows immediately from
(\ref{eq:asymptotic-iteration-formula}), the iterative upper bound
formula.
\end{thmproof}
{\it Remark:} (\ref{eq:asymptotic-iteration-formula}) is a one-dimensional
iterative bound for general asymmetric memoryless channels. In
\cite{WangKulkarniPoor04}, this iterative upper bound will be further
strengthened to:
\begin{eqnarray}
\langle CBP^{(l+1)}\rangle&\leq&\langle
CBP^{(0)}\rangle\lambda\left(1-\rho\left(1-\langle
CBP^{(l)}\rangle\right)\right),\nonumber
\end{eqnarray}
which is tight for BECs and holds for asymmetric channels as well.

\begin{thmproof}{Proof of \Theorem{\ref{thm:stability-nec}}:}
We prove this result by the erasure decomposition technique used
in~\cite{RichardsonShokrollahiUrbanke01}.

The erasure decomposition lemma in \cite{RichardsonShokrollahiUrbanke01} states
that, for any $l_0>0$, and any symmetric channel $f$ with log likelihood ratio
distribution $P^{(l_0)}$, there exists a BEC $g$ with log likelihood ratio
distribution $B^{(l_0)}$ such that $f$ is physically degraded with respect to
$g$. Furthermore, $B^{(l_0)}$ is of the following form:
\begin{eqnarray} B^{(l_0)}=2\epsilon\delta_0+(1-2\epsilon)\delta_{\infty},\nonumber
\end{eqnarray}
for all $\epsilon\leq p_e^{(l_0)}$, where $\delta_x$ is the Dirac-delta measure
centered at $x$. It can be easily shown that this erasure decomposition lemma
holds even when $f$ corresponds to a non-symmetric channel with LLR
distributions $\{P^{(l_0)}(x)\}_{x=0,1}$ and $p_e^{(l_0)}$ computed from
(\ref{eq:error-probability}).

We can then assign $B^{(l_0)}(0):=B^{(l_0)}$ and $B^{(l_0)}(1):=B^{(l_0)}\circ
I^{-1}$ to distinguish the distributions for different transmitted symbols $x$.

Suppose $r\lambda_2\rho'(1)>1$ and $\lim_{l\rightarrow \infty}p_e^{(l)}=0$. Then for any $\epsilon
>0$, $\exists l_0>0$, such that $p_e^{(l_0)}\leq\epsilon$. For simplicity, we assume
$p_e^{(l_0)}=\epsilon$. The physically better  BEC is described as above. If
during the iteration procedure (\ref{eq:main-result}), we replace the density
$P^{(l_0)}(x)$ with $B^{(l_0)}(x)$, then the resulting density will be
\begin{eqnarray}
&&\hspace{-1.2cm}P_B^{(l_0+\Delta
l)}(0)\nonumber\\
&=&2\epsilon\left(\lambda_2\rho'(1)\right)^{\Delta
l}P^{(0)}(0)\otimes
        \left(\langle P^{(0)}\rangle\right)^{\otimes
        (\Delta l-1) }\nonumber\\
    &&+\left(1-2\epsilon\left(\lambda_2\rho'(1)\right)^{\Delta l}\right)\delta_{\infty}
    + {\mathcal O}(\epsilon^2)\nonumber\\
&&\hspace{-1.2cm}P_B^{(l_0+\Delta l)}(1)\nonumber\\
&=&2\epsilon\left(\lambda_2\rho'(1)\right)^{\Delta l}P^{(0)}(1)\otimes
    \left(\langle P^{(0)}\rangle\circ I^{-1}\right)^{\otimes
    (\Delta l-1) }\nonumber\\
&&+\left(1-2\epsilon\left(\lambda_2\rho'(1)\right)^{\Delta
l}\right)\delta_{-\infty}
    + {\mathcal O}(\epsilon^2),\nonumber
\end{eqnarray}
and the averaged error probability $p_{e,B}^{(l_0+\Delta l)}$ is
\begin{eqnarray}
    p_{e,B}^{(l_0+\Delta l)}&:=&\int_{-\infty}^0\frac{P_B^{(l_0+\Delta l)}(0)+P_B^{(l_0+\Delta l)}(1)\circ
    I^{-1}}{2}(dm)\nonumber\\
        &=&
        {\mathcal O}(\epsilon^2)+2\epsilon(\lambda_2\rho'(1))^{\Delta l}\int_{-\infty}^0d\left(\langle P^{(0)}\rangle\right)^{\otimes \Delta l}.\nonumber\\ \label{eq:stability-1}
\end{eqnarray}
By the fact that $r=\langle CBP^{(0)}\rangle$ is the Chernoff bound on
$\int_{-\infty}^0d\langle P^{(0)}\rangle$, the regularity condition and the
Chernoff theorem, for any $\epsilon'>0$, there exists a large enough $\Delta l$
such that
\begin{eqnarray}
\int_{-\infty}^0d\left(\langle P^{(0)}\rangle\right)^{\otimes \Delta l}\geq (r-\epsilon')^{\Delta
l}.\nonumber
\end{eqnarray}
With a small enough $\epsilon'$, we have $\lambda_2\rho'(1)(r-\epsilon')>1$. Thus with large enough
$\Delta l$, we have
\begin{eqnarray}
p_{e,B}^{(l_0+\Delta l)}&>&{\mathcal
O}(\epsilon^2)+2\epsilon.\nonumber
\end{eqnarray}
With small enough $\epsilon$ or equivalently large enough $l_0$, we have
\begin{eqnarray}
p_{e,B}^{(l_0+\Delta l)}&>&{\mathcal
O}(\epsilon^2)+2\epsilon>\epsilon=p_e^{(l_0)}.\nonumber
\end{eqnarray}
However, by the monotonicity with respect to physically degraded channels we
have, $p_e^{(l_0+\Delta l)}\geq p_{e,B}^{(l_0+\Delta l)}>p_e^{(l_0)}$, which
contradicts the monotonicity of $p_e^{(l)}$ with respect to $l$. From the above
reasoning, if $r\lambda_2\rho'(1)>1$, then $\lim_{l\rightarrow\infty}
p_e^{(l)}> 0$, which completes the proof.
\end{thmproof}

\noindent {\it Remark:} From the sufficient stability condition,
for those codes with $\lambda_2> 0$, the convergence rate is
exponential in $l$, i.e.\
$BER=O\left((r\lambda_2\rho'(1))^l\right)$. However the number of
bits involved in the ${\mathcal N}^{2l}$ tree is
$O\left(\left((d_v-1)(d_c-1)\right)^l\right)$, which is usually
much faster than the reciprocal of the decrease rate of
$BER=O\left((r\lambda_2\rho'(1))^l\right)$. As a result, we
conjecture that the average performance of the code ensemble with
$\lambda_2>0$ will have bad {\it block} error probabilities. This
is confirmed in \Figure{\ref{fig:comparison-z}(b)} and
theoretically proved for the BEC in
\cite{OrlitskyViswanathanZhang03}. The converse is stated and
proved in the following corollary.

\begin{corollary}
Let ${\EE}\left\{ Z_{B}^{(l)}\right\}$ denote the block error probability of
codeword length $n$ after $l$ iterations of the belief propagation algorithm,
which is averaged over equiprobable codewords, channel realizations, and the
code ensemble ${\mathcal C}^n(\lambda,\rho)$. If $\lambda_2=0$ and $l_n$
satisfying $\ln\ln n=o(l_n)$ and $l_n=o(\ln n)$,
\begin{eqnarray}\lim_{n\rightarrow\infty}{\EE}\left\{Z_{B}^{(l_n)}\right\}=0.\nonumber
\end{eqnarray}
\end{corollary}
\begin{proof}
This result can be proven directly by the cycle-free convergence theorem, the
super-exponential {\it bit} convergence rate with respect to $l$, and the union
bound.
\end{proof}


A similar observation is also made and proved in \cite{JinMcEliece02}, in which
it is shown that the interleaving gain exponent of the block error rate is
$-J+2$, where $J$ is the number of parallel constituent codes. The variable
node degree $d_v$ is the number of parity check equations (parity check
sub-codes) in which a variable bit participates. In a sense, an LDPC code is
similar to $d_v$ parity check codes interleaved together. With $d_v=2$, good
interleaving gain for the block error probability is not expected.

%
%

\section{Simulations and Discussion\label{sec:simulations}}
It is worth noting that for non-symmetric channels, different codewords will
have different error-resisting capabilities. In this section, we consider the
averaged performance. We can obtain codeword-independent performance by adding
a random number to the information message before encoding and then subtracting
it after decoding. This approach, however, introduces higher computational
cost.

\subsection{Simulation Settings}
With the help of the sufficient condition of the stability theorem
(\Theorem{\ref{thm:stability-suff}}), we can use $\epsilon^*$ to set a stopping
criterion for the iterations of the density evolution. We use the 8-bit
quantized density evolution method with $(-15,15)$ being the domain of the LLR
messages. We will determine the largest thresholds such that the evolved
Chernoff bound $\langle CBP^{(l)}\rangle$ hits $\epsilon^*$ within 100
iterations, i.e.\ $\langle CBP^{(100)}\rangle<\epsilon^*$. Better performance
can be achieved by using more iterations,  which, however, is of less practical
interest. For example, the 500-iteration threshold of our best code for
z-channels, 12B (described below), is 0.2785, compared to the 100-iteration
threshold 0.2731. Five different code ensembles with rate $1/2$ are extensively
simulated, including regular $(3,6)$ codes, regular $(4,8)$ codes, 12A codes,
12B codes, and 12C codes, where
\begin{itemize}
\item 12A: 12A is a rate-$1/2$ code ensemble found by Richardson, {\it et al.}\ in \cite{RichardsonShokrollahiUrbanke01},
which is the best known degree distribution optimized for the symmetric
BiAWGNC, having maximum degree constraints $\max d_v\leq 12$ and $\max d_c\leq
9$. Its degree distributions are
\begin{eqnarray}
\lambda(x)&=&0.24426x+0.25907x^2+0.01054x^3\nonumber\\
&&+0.05510x^4+0.01455x^7+0.01275x^9\nonumber\\
&&+0.40373x^{11},\nonumber\\
\rho(x)&=&0.25475x^6+0.73438x^7+0.01087x^8.\nonumber
\end{eqnarray}
\item 12B: 12B is a rate-$1/2$ code ensemble obtained by minimizing the hitting time of $\epsilon^*$ in z-channels, through hill-climbing and
linear programming techniques. The maximum degree constraints are also $\max
d_v\leq 12$ and $\max d_c\leq 9$. The differences between 12A and 12B are (1)
12B is optimized for the z-channels with our codeword-averaged density
evolution, and 12A is optimized for the symmetric BiAWGNC. (2) 12B is optimized
with respect to the hitting time of $\epsilon^*$ (depending on
$(\lambda,\rho)$) rather than a fixed small threshold. The degree distributions
of 12B are
\begin{eqnarray}
\lambda(x)&=&0.236809x+0.309590x^2+0.032789x^3\nonumber\\
&&+0.007116x^4+0.000001x^5+ 0.413695x^{11},\nonumber\\
\rho(x)&=&0.000015x^5+0.464854x^6+0.502485x^7\nonumber\\
&&+0.032647x^8.\nonumber
\end{eqnarray}
\item 12C: 12C a rate-$1/2$ code ensemble similar to 12B, but with $\lambda_2$ being hard-wired to $0$, which is suggested by the convergence rate
in the sufficient stability condition. The degree distributions of 12C are
\begin{eqnarray}
\lambda(x)&=&0.861939x^2+0.000818x^3+0.000818x^4\nonumber\\
&&+0.000818x^5+0.000818x^6+0.000818x^7\nonumber\\
&&+0.000218x^8+0.077898x^9+0.055843x^{10}\nonumber\\
&&+0.000013x^{11},\nonumber\\
\rho(x)&=&0.000814x^4+0.560594x^5+0.192771x^6\nonumber\\
&&+0.145207x^7+0.100613x^8.\nonumber
\end{eqnarray}
\end{itemize}

\begin{table*}
\centering\footnotesize
\begin{tabular}{cccccccccc}
\hline\hline
\multirow{2}{1.5cm}{Codes}&\multicolumn{2}{c}{BEC}&\multicolumn{2}{c}{BSC}&\multicolumn{2}{c}{Z-channels}&\multicolumn{2}{c}{BiAWGNC}&\multirow{2}{1.5cm}{Stability
{\tiny$\langle CBP\rangle$}}\\
 \cline{2-9}
 &$\epsilon$&{\tiny $\langle CBP\rangle$}& $\epsilon$ &{\tiny $\langle CBP\rangle$} & $\epsilon_1$& {\tiny $\langle CBP\rangle$} & $\sigma$& {\tiny $\langle CBP\rangle$} & \\
 \hline
 (3,6)&0.4294&0.4294&0.0837&0.5539&0.2305&0.4828&0.8790&0.5235&--\\
 (4,8)&0.3834&0.3834&0.0764&0.5313&0.1997&0.4497&0.8360&0.4890&--\\
 12A &0.4682 & 0.4682 & 0.0937 & 0.5828 & 0.2710 & 0.5233 & 0.9384 & 0.5668 & 0.6060 \\
 12B &0.4753 & 0.4753 & 0.0939 & 0.5834 & 0.2731 & 0.5253 & 0.9362 & 0.5653 & 0.6247 \\
 12C & 0.4354 & 0.4354 & 0.0862 & 0.5613 & 0.2356 & 0.4881 & 0.8878 & 0.5303 & -- \\
 Sym.\ Info.\ Rate &0.5000 & 0.5000&0.1100&0.6258&0.2932& 0.5415 &0.9787&0.5933&--\\
 Capacity &0.5000 & 0.5000&0.1100&0.6258&0.3035& 0.5509 &0.9787&0.5933&--\\
 \hline
 \hline
\end{tabular}
\caption{Thresholds of different codes and channels, with precision
$10^{-4}$.}\label{tab:threshold-comparison}
\end{table*}
Four different channels are considered, including the BEC, BSC, z-channel, and
BiAWGNC. Z-channels are simulated by  binary non-symmetric channels with very
small $\epsilon_0$ ($\epsilon_0=0.00001$) and different values of $\epsilon_1$.
\Table{\ref{tab:threshold-comparison}} summarizes the thresholds with precision
$10^{-4}$. Thresholds are not only presented by their conventional channel
parameters, but also by their Bhattacharyya noise parameters (Chernoff bounds).
The column ``stability" lists the maximum $r:=\langle CBP^{(0)}\rangle$ such
that $r\lambda_2\rho'(1)<1$, which is an upper bound on the $\langle
CBP^{(0)}\rangle$ values of decodable channels.  Further discussion of the
relationship between $\langle CBP^{(0)}\rangle$ and the decodable threshold can
be found in~\cite{WangKulkarniPoor04}.

From \Table{\ref{tab:threshold-comparison}}, we observe that 12A outperforms
12B in Gaussian channels (for which 12A is optimized), but 12B is superior in
z-channels for  which it is optimized. The above behavior promises room for
improvement with codes optimized for different channels, as was also shown
in~\cite{HouSiegelMilistein01}.

\begin{figure}[t]
\centering \includegraphics[width=8.5cm,
keepaspectratio=true]{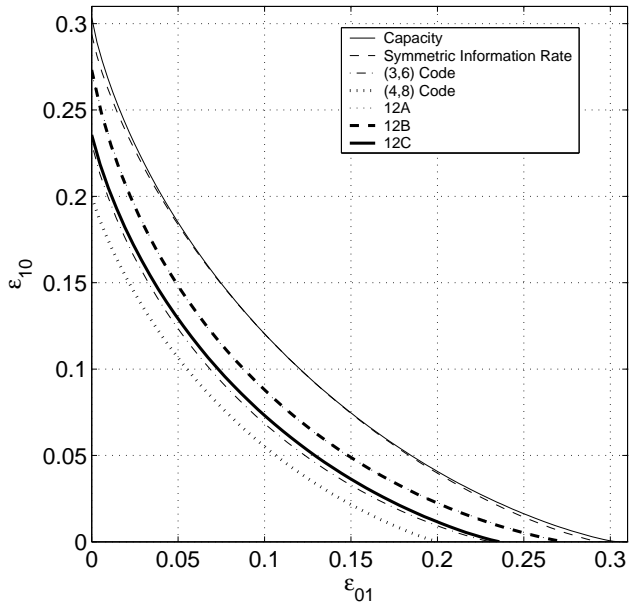}\caption{Asymptotic thresholds  and the
achievable regions of different codes in binary asymmetric channels. }
\label{fig:bc-capacity-1}
\end{figure}
\Figure{\ref{fig:bc-capacity-1}} demonstrates the asymptotic thresholds of
these codes in binary asymmetric channels (BASCs) with the curves of 12A and
12B being very close together. It is seen that 12B is slightly better when
$\epsilon_0,\epsilon_1\rightarrow 0$ or $\epsilon_0\approx\epsilon_1$. We
notice that all the achievable regions of these codes are bounded by the
symmetric mutual information rate (with a $(1/2,1/2)$ {\it a priori}
distribution), which was also suggested in \cite{KavcicMaMitzanmacher03}. The
difference between the symmetric mutual information rate and the capacity for
non-symmetric channels is generally indistinguishable from the practical point
of view. For example, in \cite{MajaniRumsey91}, it was shown that the ratio
between the symmetric mutual information rate and the capacity is lower bounded
by $\frac{e\ln 2}{2}\approx0.942$. \cite{ShulmanFeder04} further proved that
the absolute difference is upper bounded by $0.011$ bit/sym. Further discussion
of capacity achieving codes with non-uniform {\it a priori} distributions can
be found in \cite{McEliece01a} and \cite{BennatanBurshtein04}.


\begin{figure*}[t]
\centering
\begin{tabular}{cc}
\includegraphics[width=8.5cm,
keepaspectratio=true]{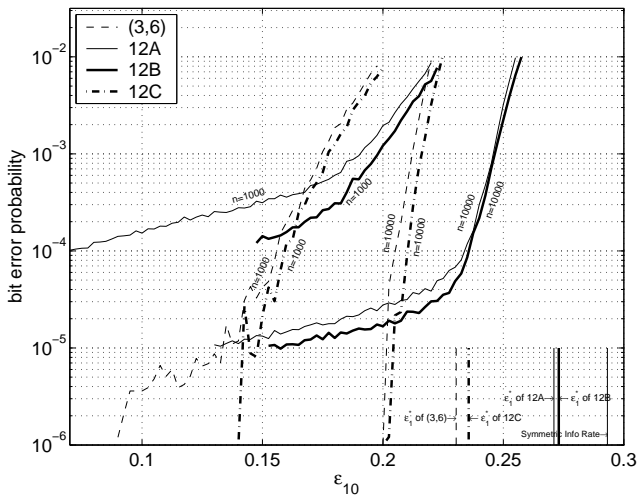}&\includegraphics[width=8.5cm,
keepaspectratio=true]{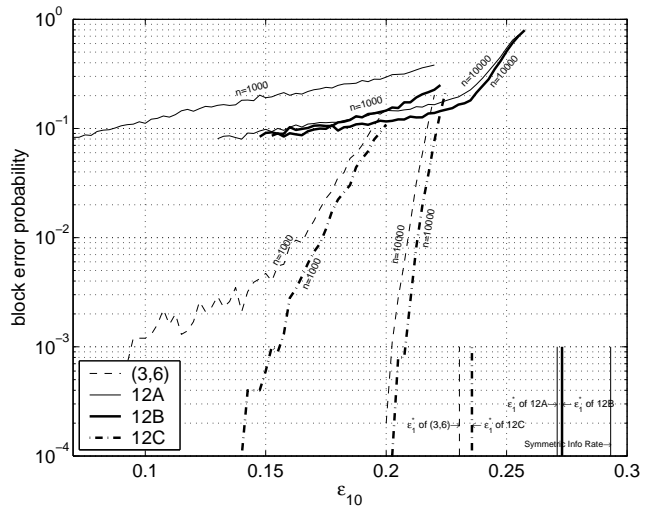}\\
(a) Bit error rates& (b) Block error rates
\end{tabular}
\caption{Bit/block error rates versus $\epsilon_1$ with fixed
$\epsilon_0=0.00001$. Computed thresholds for symmetric mutual information
rate, (3,6), 12A, 12B, and 12C codes are $0.2932$, $0.2305$, $0.2710$,
$0.2730$, and $0.2356$, respectively. $40$ iterations of belief propagation
algorithms were performed. 10,000 codewords were used for the simulations.}
\label{fig:comparison-z}
\end{figure*}
\Figures{\ref{fig:comparison-z}(a) and \ref{fig:comparison-z}(b)} consider
several fixed finite codes in z-channels. We arbitrarily select graphs from the
code ensemble with codeword lengths $n=$1,000 and $n=$10,000. Then, with these
graphs (codes) fixed, we find the corresponding parity matrix $\mathbf A$, use
Gaussian elimination to find the generator matrix $\mathbf G$, and  transmit
different codewords by encoding equiprobably selected information messages.
Belief propagation decoding is used with $40$ iterations for each codeword.
10,000 codewords are transmitted, and the overall bit/block error rates versus
different $\epsilon_1$ are plotted for different code ensembles and codeword
lengths. Our new density evolution predicts the waterfall region quite
accurately when the bit error rates are of primary interest. Though there are
still gaps between the performance of finite codes and our asymptotic
thresholds, the performance gaps between different finite length codes are very
well predicted by the differences between their asymptotic thresholds. From the
above observations and the underpinning theorems, we see that our new density
evolution is a successful generalization of the traditional one from both
practical and theoretical points of view.

\Figure{\ref{fig:comparison-z}(b)} exhibits the block error rate of the same
10,000-codeword simulation. The conjecture of bad block error probabilities for
$\lambda_2>0$ codes is confirmed. Besides the conjectured bad block error
probabilities, \Figures{\ref{fig:comparison-z}(a) {\rm
and}~\ref{fig:comparison-z}(b)} also suggest that codes with $\lambda_2=0$ will
have a better error floor compared to those with $\lambda_2>0$, which can be
partly explained by the comparatively slow convergence speed stated in the
sufficient stability condition for $\lambda_2>0$ codes. 12C is so far the best
code we have for $\lambda_2=0$. However, its threshold is not as good as those
of 12A and 12B. If good block error rate and low error floor are our major
concerns, 12C (or other codes with $\lambda_2=0$) can still be competitive
choices. Recent results in \cite{YangRyan03} shows that the error floor for
codes with $\lambda_2>0$ can be lowered by carefully arranging the degree two
variable nodes in the corresponding graph while keeping a similar waterfall
threshold.

\begin{figure*}[t]
\centering \begin{tabular}{cc}
\includegraphics[width=8.5cm,
keepaspectratio=true]{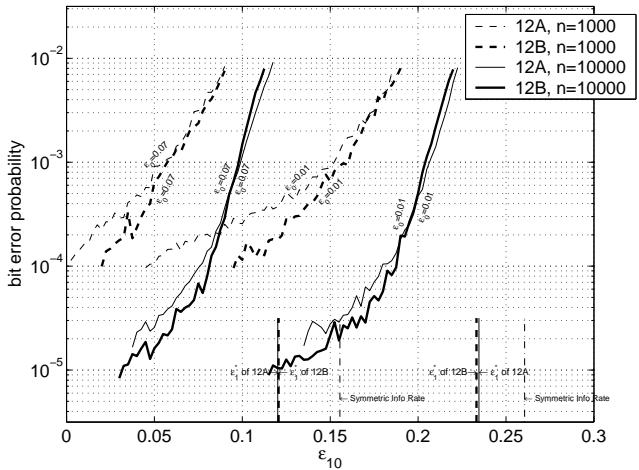} &
\includegraphics[width=8.5cm,
keepaspectratio=true]{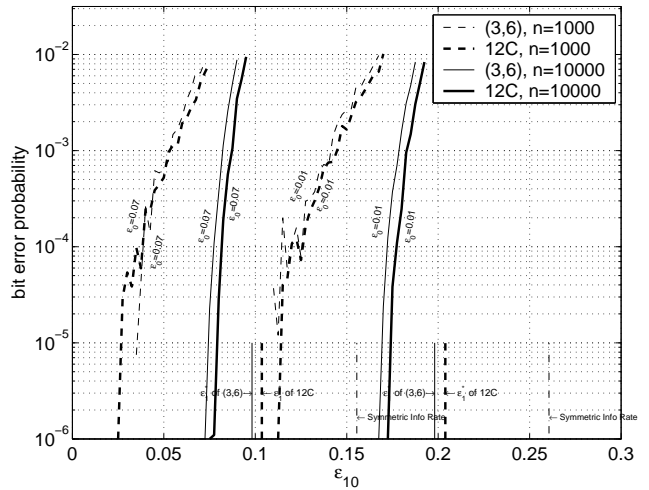}\\
(a) 12A \& 12B: & (b) 12C \& regular (3,6) codes
\end{tabular}
\caption{Bit error rates versus $\epsilon_1$ for $\epsilon_0=0.01$ and
$\epsilon_0=0.7$. The DE thresholds of (12A, 12B, 12C, (3,6)) are
$(0.2346,0.2332,0.2039, 0.1981)$ for $\epsilon_0=0.01$ and
$(0.1202,0.1206,0.1036,0.0982)$ for $\epsilon_0=0.07$.
  $40$ iterations
of belief propagation algorithms were performed. 2,000 codewords were used for
the simulations.} \label{fig:dv12-2DO-predictability}
\end{figure*}
\Figures{\ref{fig:dv12-2DO-predictability}(a) and
\ref{fig:dv12-2DO-predictability}(b)} illustrate the bit error rates versus
different BASC settings with 2,000 transmitted codewords. Our computed density
evolution threshold is again highly correlated with the performance of finite
length codes for different asymmetric channel settings.

We close this section by highlighting two applications of our results.
\begin{enumerate}
\item Error Floor Analysis: ``The error floor" is a characteristic of
iterative decoding algorithms, which is of practical importance and may not be
able to be determined solely by simulations. More analytical tools are needed
to find error floors for corresponding codes. Our convergence rate statements
in the sufficient stability condition may shed some light on finding codes with
low error floors.
\item Capacity-Approaching Codes for General Non-Standard
Channels: Various {\it very good}  codes (capacity-approaching) are known for
standard channels, but very good codes for non-standard channels are not yet
known. It is well known that  one can construct capacity-approaching codes by
incorporating symmetric-information-rate-approaching linear codes with the
symbol mapper and demapper as an inner code
\cite{BennatanBurshtein04,McEliece01a,RatzerMacKay03}. Understanding density
evolution for general memoryless channels allows us  to construct such
symmetric-information-rate-approaching codes (for non-symmetric memoryless
channels), and thus to find capacity-approaching codes after concatenating the
inner symbol mapper and demapper. It is worth noting that intersymbol
interference channels are dealt with by Kav\v{c}i\'{c} {\it et al.}\ in
\cite{KavcicMaMitzanmacher03} using the coset codes approach. It will be of
great help if a unified framework for non-symmetric channels with memory can be
found by incorporating both coset codes and codeword averaging approaches.
\end{enumerate}

\section{Further Implications of Generalized Density Evolution\label{sec:side-results}}
\subsection{Typicality of Linear LDPC Codes}
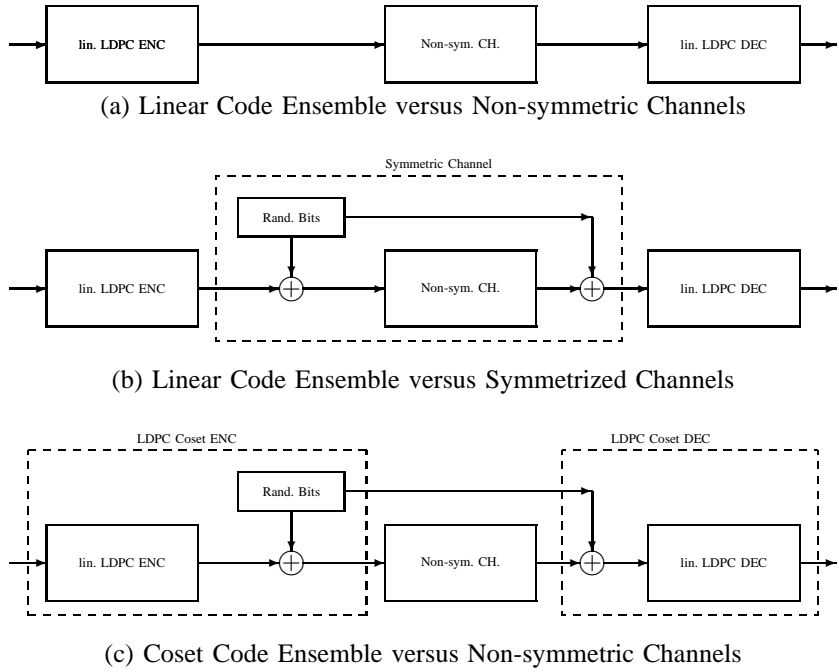
\begin{figure*}
\centering \setlength{\unitlength}{.1mm}
\begin{tabular}{c}
\begin{picture}(1000,100)

\put(-50,50){\vector(1,0){50}}
\put(0,0){\framebox(200,100){\tiny lin.\ LDPC ENC}}
%
\put(200,50){\vector(1,0){250}}
\put(450,0){\framebox(200,100){\tiny Non-sym.\ CH.}}
%
\put(650,50){\vector(1,0){150}}
\put(1000,50){\vector(1,0){50}}
 \put(800,0){\framebox(200,100){\tiny lin.\ LDPC
DEC}}
\put(-50,50){\vector(1,0){50}}
\put(0,0){\framebox(200,100){\tiny lin.\ LDPC ENC}}

\end{picture}\\
(a) Linear Code Ensemble versus Non-symmetric Channels\\
\begin{picture}(1000,310)(0,-40)

 \put(-50,50){\vector(1,0){50}}
\put(0,0){\framebox(200,100){\tiny lin.\ LDPC ENC}}
\put(325,50){\picchk}
%
\put(255,120){\framebox(140,50){\tiny Rand.\ Bits}}
\put(325,120){\vector(0,-1){55}}
\put(395, 145){\vector(1,0){330}}
\put(725,145){\vector(0,-1){80}}
%
\put(200,50){\vector(1,0){110}}
\put(340,50){\vector(1,0){110}}
\put(450,0){\framebox(200,100){\tiny Non-sym.\ CH.}}
%
%
\put(650,50){\vector(1,0){60}}
%
\put(725,50){\picchk}
\put(740,50){\vector(1,0){60}}
%
\put(1000,50){\vector(1,0){50}}
 \put(800,0){\framebox(200,100){\tiny lin.\ LDPC
DEC}}

\put(225,-20){\dashbox{10}(540,220){}} \put(450,210){\tiny Symmetric Channel}

\end{picture}
\\(b) Linear Code Ensemble versus Symmetrized Channels\\
\begin{picture}(1000,310)(0,-40)

 \put(-50,50){\vector(1,0){50}}
\put(0,0){\framebox(200,100){\tiny lin.\ LDPC ENC}}
\put(325,50){\picchk}
%
\put(255,120){\framebox(140,50){\tiny Rand.\ Bits}}
\put(325,120){\vector(0,-1){55}}
\put(395, 145){\vector(1,0){330}}
\put(725,145){\vector(0,-1){80}}
%
\put(200,50){\vector(1,0){110}}
\put(340,50){\vector(1,0){110}}
\put(450,0){\framebox(200,100){\tiny Non-sym.\ CH.}}
%
%
\put(650,50){\vector(1,0){60}}
%
\put(725,50){\picchk}
\put(740,50){\vector(1,0){60}}
%
\put(1000,50){\vector(1,0){50}}
 \put(800,0){\framebox(200,100){\tiny lin.\ LDPC
DEC}}

 \put(-25,-20){\dashbox{10}(450,220){}} \put(120,210){\tiny LDPC
Coset ENC} \put(685,-20){\dashbox{10}(340,220){}} \put(750,210){\tiny LDPC
Coset DEC}

\end{picture}
\\ (c) Coset Code Ensemble versus Non-symmetric Channels
\end{tabular}
\caption{Comparison of the approaches based on codeword averaging and the coset
code ensemble.\label{fig:coset-code-arg}}
\end{figure*}

One reason that non-symmetric channels are often overlooked is we
can always transform a non-symmetric channel into a symmetric
channel. Depending on different points of view, this
channel-symmetrizing technique is termed the coset code argument
\cite{KavcicMaMitzanmacher03} or dithering/the i.i.d.\ channel
adapter \cite{HouSiegelMilisteinPfister03}, as illustrated in
\Figures{\ref{fig:coset-code-arg}(c) {\rm
and}~\ref{fig:coset-code-arg}(b)}. Our generalized density
evolution provides a simple way to directly analyze the linear
LDPC code ensemble on non-symmetric channels, as in
\Figure{\ref{fig:coset-code-arg}(a)}.

As shown in \Theorems{\ref{thm:stability-suff} {\rm
and}~\ref{thm:stability-nec}}, the necessary and sufficient
stability conditions of linear LDPC codes for non-symmetric
channels, \Figure{\ref{fig:coset-code-arg}(a)}, are identical to
those of the coset code ensemble,
\Figure{\ref{fig:coset-code-arg}(c)}. Monte Carlo simulations
based on finite-length codes ($n =10^4$)
\cite{HouSiegelMilisteinPfister03} further show that the
codeword-averaged performance in
\Figure{\ref{fig:coset-code-arg}(a)} is nearly
identical\footnote{That is, it is within the precision of the
Monte Carlo simulation.} to the performance of
\Figure{\ref{fig:coset-code-arg}(c)} when the same encoder/decoder
pair is used. The above two facts suggest a close relationship
between linear codes and the coset code ensemble, and it was
conjectured in \cite{HouSiegelMilisteinPfister03} that the scheme
in \Figure{\ref{fig:coset-code-arg}(a)} should always have the
same/similar performance as those illustrated by
\Figure{\ref{fig:coset-code-arg}(c)}. This short subsection is
devoted to the question whether the systems in
\Figures{\ref{fig:coset-code-arg}(a) {\rm
and}~\ref{fig:coset-code-arg}(c)} are equivalent in terms of
performance. In sum, the performance of the linear code ensemble
is very unlikely to be identical to that of the coset code
ensemble. However, when the minimum
$d_{c,min}:=\{k\in\NN:\rho_k>0\}$ is sufficiently large, we can
prove that their performance discrepancy is theoretically
indistinguishable. In practice, the discrepancy for $d_{c,min}\geq
6$ is $<0.05\%$.

\begin{figure}
\centering {\includegraphics[width=8.5cm, keepaspectratio=true]{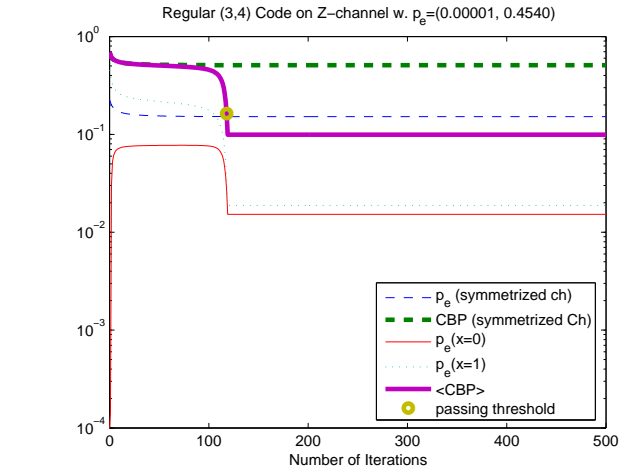}}
\caption{Density evolution for z-channels with the linear code ensemble and the
coset code ensemble.\label{fig:dens-evo-path}}
\end{figure}

Let $P_{a.p.}^{(l)}(0):=P^{(l)}(0)$ and $P_{a.p.}^{(l)}(1):=P^{(l)}(1)\circ
I^{-1}$ denote the two evolved densities with {\it aligned parity}, and
similarly define $Q_{a.p.}^{(l)}(0):=Q^{(l)}(0)$ and
$Q_{a.p.}^{(l)}(1):=Q^{(l)}(1)\circ I^{-1}$. Our main result in
(\ref{eq:main-result}) can be rewritten in the following form:
 \begin{eqnarray}
    P^{(l)}_{a.p.}(x)&=&P^{(0)}_{a.p.}(x)\otimes
        \lambda\left(Q^{(l-1)}_{a.p.}(x)\right)\nonumber\\
    Q^{(l-1)}_{a.p.}(x)&=&
         \Gamma^{-1}\left(\rho\left(\Gamma\left(\frac{P^{(l-1)}_{a.p.}(0)+P^{(l-1)}_{a.p.}(1)}{2}\right)\right)\right.\nonumber\\
     &&\left.       +(-1)^x\rho\left(\Gamma\left(\frac{P^{(l-1)}_{a.p.}(0)-P^{(l-1)}_{a.p.}(1)}{2}\right)\right)
        \right).\nonumber\\ \label{eq:dens-linear}
    \end{eqnarray}
Let $p^{(l)}_{e,linear}$ denote the corresponding bit error probability of the
linear codes after $l$ iterations. For comparison, the traditional formula of
density evolution for the  symmetrized channel (the coset code ensemble) is as
follows:
 \begin{eqnarray}
    P^{(l)}_{coset}&=&P^{(0)}_{coset}\otimes
        \lambda\left(Q^{(l-1)}_{coset}\right)\nonumber\\
    Q^{(l-1)}_{coset}&=&
         \Gamma^{-1}\left(\rho\left(\Gamma\left(P^{(l-1)}_{coset}\right)\right)
        \right),\label{eq:dens-coset}
    \end{eqnarray}
where $P^{(0)}_{coset}=\frac{\sum_{x=0,1}P^{(0)}_{a.p.}(x)}{2}$. Similarly, let
$p^{(l)}_{e,coset}$ denote the corresponding bit error probability.

It is clear from the above formulae that when the channel of interest is
symmetric, namely $P^{(0)}_{a.p.}(0)=P^{(0)}_{a.p.}(1)$, then
$P^{(l)}_{coset}=P^{(l)}_{a.p.}(0)=P^{(l)}_{a.p.}(1)$ for all $l\in\NN$.
However, for non-symmetric channels, since the variable node iteration involves
convolution of several densities given the same $x$ value, the difference
between $Q^{(l-1)}_{a.p.}(0)$ and $Q^{(l-1)}_{a.p.}(1)$ will be amplified after
each variable node iteration. Hence it is very unlikely that the decodable
thresholds of linear codes and coset codes will be analytically identical,
namely
\begin{eqnarray}
\lim_{l\rightarrow \infty}p^{(l)}_{e,linear}=0
\stackrel{?}{\Longleftrightarrow} \lim_{l\rightarrow
\infty}p^{(l)}_{e,coset}=0.\nonumber
\end{eqnarray}
\Figure{\ref{fig:dens-evo-path}} demonstrates the traces of the evolved
densities for the regular (3,4) code on z-channels. With the one-way crossover
probability being 0.4540, the generalized density evolution for linear codes is
able to converge within 179 iterations, while the coset code ensemble shows no
convergence within 500 iterations. This demonstrates the possible performance
discrepancy, though we do not have analytical results proving that the latter
will not converge after further iterations. \Table{\ref{tab:typicality}}
compares the decodable thresholds such that the density evolution enters the
stability region within 100 iterations. We notice that the larger $d_{c,min}$
is, the smaller the discrepancy is. This phenomenon can be characterized by the
following theorem.
\begin{table}
\centering \caption{Threshold  comparison $p^*_{1\rightarrow 0}$ of linear and
coset LDPC codes on Z-channels\label{tab:typicality}}
\begin{tabular}{l|c|c}
\hline \hline
 $(\lambda,\rho)$       & $(x^2,x^3)$   & $(x^2,x^5)$   \\
 \hline
Linear                  & 0.4540        & 0.2305        \\
Coset                   & 0.4527        & 0.2304        \\
\hline \hline \multicolumn{3}{c}{~}\\
\multicolumn{3}{c}{~}\\
\hline \hline
 $(\lambda,\rho)$       & $(x^2,0.5x^2+0.5x^3)$ & $(x^2,0.5x^4+0.5x^5)$\\
 \hline
Linear                  & 0.5888                & 0.2689\\
Coset                   & 0.5908                & 0.2690\\
\hline \hline
\end{tabular}
\end{table}

\begin{theorem}\label{thm:typicality}
Consider non-symmetric memoryless channels and a fixed pair of finite-degree
polynomials $\lambda$ and $\rho$. The shifted version of the check node
polynomial is denoted as $\rho_{\Delta}=x^{\Delta}\cdot\rho$ where
$\Delta\in\NN$. Let $P^{(l)}_{coset}$ denote the evolved density from the coset
code ensemble with degrees $(\lambda,\rho_{\Delta})$, and $\langle
P^{(l)}\rangle=\frac{1}{2}\sum_{x=0,1}P^{(l)}_{a.p.}(x)$ denote the averaged
density from the linear code ensemble with degrees $(\lambda,\rho_{\Delta})$.
For any $l_0\in\NN$, $\lim_{\Delta\rightarrow\infty}\langle
P^{(l)}\rangle\stackrel{\mathcal D}{=}P^{(l)}_{coset}$ in distribution for all
$l\leq l_0$, with the convergence rate for each iteration being
$\bigorder\left(\const^{\Delta}\right)$ for some $\const<1$.
\end{theorem}
\begin{corollary}[The Typicality Results for Z-Channels]\label{cor:typi-z} For any $\epsilon > 0$, there exists a $\Delta\in\NN$ such that
{\footnotesize \begin{eqnarray} \left|\sup\left\{p_{1\rightarrow
0}:\lim_{l\rightarrow\infty}p^{(l)}_{e,linear}=0\right\}-\sup\left\{p_{1\rightarrow
0}:\lim_{l\rightarrow\infty}p^{(l)}_{e,coset}=0\right\}\right|\nonumber\\
<\epsilon.\nonumber
\end{eqnarray}}\cmpt
Namely, the asymptotic decodable thresholds of the linear and the coset code
ensemble are arbitrarily close when the minimum check node degree $d_{c,min}$
is sufficiently large.
\end{corollary}

Similar corollaries can be constructed for other channel models
with different types of noise parameters, e.g., the $\sigma^*$ in
the composite BiAWGNC. A proof of \Corollary{\ref{cor:typi-z}} is
found in \Append{\ref{app:proof-typicality-z}}.

\begin{thmproof}{Proof of \Theorem{\ref{thm:typicality}}:}
Since the functionals in (\ref{eq:dens-linear}) and (\ref{eq:dens-coset}) are
continuous with respect to convergence in distribution, we need only to show
that $\forall l\in\NN$,
\begin{eqnarray}
&&\lim_{\Delta\rightarrow\infty}Q^{(l-1)}_{a.p.}(0)\stackrel{\mathcal D}{=}\lim_{\Delta\rightarrow\infty}Q^{(l-1)}_{a.p.}(1)\nonumber\\
&&\stackrel{\mathcal
D}{=}~\Gamma^{-1}\left(\rho\left(\Gamma\left(\frac{P^{(l-1)}_{a.p.}(0)+P^{(l-1)}_{a.p.}(1)}{2}\right)\right)\right)\nonumber\\
&&=~\frac{Q^{(l-1)}_{a.p.}(0)+Q^{(l-1)}_{a.p.}(1)}{2}
,\label{eq:typicality-temp-1}
\end{eqnarray}
where $\stackrel{\mathcal D}{=}$ denotes convergence in distribution. Then by
inductively applying this weak convergence argument, for any bounded $l_0$,
$\lim_{\Delta\rightarrow\infty}\langle P^{(l)}\rangle\stackrel{\mathcal
D}{=}P^{(l)}_{coset}$ in distribution for all $l\leq l_0$. Without loss of
generality,\footnote{We also need to assume that $\forall x,
P^{(l-1)}_{a.p.}(x)(m=0)=0$ so that $\ln\coth\left|\frac{m}{2}\right|\in\RR^+$
almost surely. This assumption can be relaxed by separately considering the
event that $m_{in,i}=0$ for some $i\in\{1,\cdots,d_c-1\}$.} we may assume
$\rho_{\Delta}=x^\Delta$ and prove the weak convergence of distributions on the
domain
\begin{eqnarray}
\gamma(m)&:=&\left(1_{\{m\leq0\}},\ln\coth\left|\frac{m}{2}\right|\right)\nonumber\\
&=&(\gamma_1,\gamma_2)\in\GF(2)\times \RR^+,\nonumber
\end{eqnarray}
on which the check node iteration becomes
\begin{eqnarray}
\gamma_{out,\Delta}=\gamma_{in,1}+\gamma_{in,2}+\cdots +
\gamma_{in,\Delta}.\nonumber
\end{eqnarray}
Let $P'_0$ denote the density of $\gamma_{in}(m)$ given that the distribution
of $m$ is $P^{(l-1)}_{a.p.}(0)$ and let $P'_1$ similarly correspond to
$P^{(l-1)}_{a.p.}(1)$. Similarly let $Q'_{0,\Delta}$ and $Q'_{1,\Delta}$ denote
the output distributions on $\gamma_{out,\Delta}$ when the check node degree is
$\Delta+1$. It is worth noting that any pair of $Q'_{0,\Delta}$ and
$Q'_{1,\Delta}$ can be mapped bijectively to the LLR distributions
$Q^{(l-1)}_{a.p.}(0)$ and $Q^{(l-1)}_{a.p.}(1)$.

 Let
$\Phi_{P'}(k,r):=\EE_{P'}\left\{(-1)^{k\gamma_1}e^{ir\gamma_2}\right\}, \forall
k\in\NN, r\in\RR$, denote the Fourier transform of the density $P'$. Proving
(\ref{eq:typicality-temp-1}) is equivalent to showing that
\begin{eqnarray}
\forall k\in\NN, r\in\RR,~
\lim_{\Delta\rightarrow\infty}\Phi_{Q'_{0,\Delta}}(k,{r})=
\lim_{\Delta\rightarrow\infty}\Phi_{Q'_{1,\Delta}}(k,{r}).\nonumber
\end{eqnarray}
However, to deal with the strictly growing average of the ``limit
distribution", we concentrate on the distribution of the normalized output
$\frac{\gamma_{out,\Delta}}{\Delta}$ instead. We then need to prove that
\begin{eqnarray}
\forall k\in\NN, r\in\RR,~
\lim_{\Delta\rightarrow\infty}\Phi_{Q'_{0,\Delta}}(k,\frac{r}{\Delta})=
\lim_{\Delta\rightarrow\infty}\Phi_{Q'_{1,\Delta}}(k,\frac{r}{\Delta}).\nonumber
\end{eqnarray}
We first note that for all $x=0,1$, $Q'_{x,\Delta}$ is the averaged
distribution of $\gamma_{out,\Delta}$ when the inputs $\gamma_{in,i}$ are
governed by $P^{(l)}_{a.p.}(x_i)$ satisfying $\sum_{i=1}^\Delta x_i=x$. From
this observation, we can derive the  following iterative equations:
$\forall\Delta\in\NN$, {\small
\begin{eqnarray}
&&\Phi_{Q'_{0,\Delta}}(k,\frac{r}{\Delta})\nonumber\\
&&=\frac{\Phi_{Q'_{0,\Delta-1}}(k,\frac{r}{\Delta})\Phi_{P'_0}(k,\frac{r}{\Delta})+\Phi_{Q'_{1,\Delta-1}}(k,\frac{r}{\Delta})\Phi_{P'_1}(k,\frac{r}{\Delta})}{2}\nonumber\\
&&\Phi_{Q'_{1,\Delta}}(k,\frac{r}{\Delta})\nonumber\\
&&=\frac{\Phi_{Q'_{0,\Delta-1}}(k,\frac{r}{\Delta})\Phi_{P'_1}(k,\frac{r}{\Delta})+\Phi_{Q'_{1,\Delta-1}}(k,\frac{r}{\Delta})\Phi_{P'_0}(k,\frac{r}{\Delta})}{2}.\nonumber
\end{eqnarray}}\cmpt
 By induction, the difference thus becomes
\begin{eqnarray}
&&\hspace{-.7cm}\Phi_{Q'_{0,\Delta}}(k,\frac{r}{\Delta})-\Phi_{Q'_{1,\Delta}}(k,\frac{r}{\Delta})\nonumber\\
&=&\left(\Phi_{Q'_{0,\Delta-1}}(k,\frac{r}{\Delta})-\Phi_{Q'_{1,\Delta-1}}(k,\frac{r}{\Delta})\right)\nonumber\\
&&\cdot
        \left(\frac{\Phi_{P'_0}(k,\frac{r}{\Delta})-\Phi_{P'_1}(k,\frac{r}{\Delta})}{2}\right)\nonumber\\
        &=&2\left(\frac{\Phi_{P'_0}(k,\frac{r}{\Delta})-\Phi_{P'_1}(k,\frac{r}{\Delta})}{2}\right)^\Delta.\label{eq:typicality-temp2}
\end{eqnarray}
By Taylor's expansion and the BASC decomposition argument
in~\cite{WangKulkarniPoor04}, we can show that for all $k\in\NN$,
$r\in\RR$, and for all possible $P'_0$ and $P'_1$, the quantity in
(\ref{eq:typicality-temp2}) converges to zero with convergence
rate $\bigorder\left(\const^\Delta\right)$ for some $\const<1$. A
detailed derivation of the convergence rate is given in
\Append{\ref{app:conv-rate}}. Since the limit of the right-hand
side of (\ref{eq:typicality-temp2}) is zero, the proof of weak
convergence is complete. The exponentially fast convergence rate
$\bigorder\left(\const^\Delta\right)$ also justifies the fact that
even for moderate $d_{c,min}\geq 6$, the performances of linear
and coset LDPC codes are very close.
\end{thmproof}

{\it Remark 1:} Consider any non-perfect message distribution, namely, $\exists
    x_0$ such that $P^{(l-1)}_{a.p.}(x_0)\neq \delta_{\infty}$.    A persistent reader may notice that
$\forall x, \lim_{\Delta\rightarrow\infty}Q^{(l-1)}_{a.p.}(x)\stackrel{\mathcal
D}{=}\delta_0$, namely, as $\Delta$ becomes large, all information is erased
after passing a check node of large degree. If this convergence (erasure
effect) occurs earlier than the convergence of $Q^{(l-1)}_{a.p.}(0)$ and
$Q^{(l-1)}_{a.p.}(1)$, the performances of linear and coset LDPC codes are
``close" only when the code is ``useless."\footnote{To be more precise, it
corresponds to an extremely high-rate code and the information is erased after
every check node iteration.} To quantify the convergence rate, we consider
again the distributions on $\gamma$ and their Fourier transforms. For the
average of the output distributions $Q^{(l-1)}_{a.p.}(x)$, we have
\begin{eqnarray}
&&\hspace{-.7cm}\frac{\Phi_{Q'_{0,\Delta}}(k,\frac{r}{\Delta})+\Phi_{Q'_{1,\Delta}}(k,\frac{r}{\Delta})}{2}\nonumber\\
&=&\left(\frac{\Phi_{Q'_{0,\Delta-1}}(k,\frac{r}{\Delta})+\Phi_{Q'_{1,\Delta-1}}(k,\frac{r}{\Delta})}{2}\right)\nonumber\\
&&\cdot        \left(\frac{\Phi_{P'_0}(k,\frac{r}{\Delta})+\Phi_{P'_1}(k,\frac{r}{\Delta})}{2}\right)\nonumber\\
        &=&\left(\frac{\Phi_{P'_0}(k,\frac{r}{\Delta})+\Phi_{P'_1}(k,\frac{r}{\Delta})}{2}\right)^\Delta.\label{eq:typicality-temp3}
\end{eqnarray}
By Taylor's expansion and the BASC decomposition argument, one can show that
the limit of (\ref{eq:typicality-temp3}) exists and the convergence rate is
$\bigorder(\Delta^{-1})$. (A detailed derivation is included in
\Append{\ref{app:conv-rate}}.) This convergence rate is much slower than the
exponential rate $\bigorder\left(\const^\Delta\right)$ in the proof of
\Theorem{\ref{thm:typicality}}. Therefore, we do not need to worry about the
case in which the required $\Delta$ for the convergence of
$Q^{(l-1)}_{a.p.}(0)$ and $Q^{(l-1)}_{a.p.}(1)$ is excessively large so that
$\forall x\in\GF(2), Q^{(l-1)}_{a.p.}(x)\stackrel{\mathcal
D}{\approx}\delta_0$.

{\it Remark 2:} The intuition behind
\Theorem{\ref{thm:typicality}} is that when the minimum $d_c$ is
sufficiently large, the parity check constraint becomes relatively
less stringent. Thus we can approximate the density of the
outgoing messages for linear codes by assuming all bits involved
in that particular parity check equation are ``independently"
distributed among $\{0,1\}$, which leads to the formula for the
coset code ensemble. On the other hand, extremely large $d_c$ is
required for a check node iteration to completely destroy all
information coming from the previous iteration. This explains the
difference between their convergence rates:
$\bigorder\left(\const^\Delta\right)$ versus
$\bigorder(\Delta^{-1})$.

\begin{figure}
\centering {\includegraphics[width=8.5cm,
keepaspectratio=true]{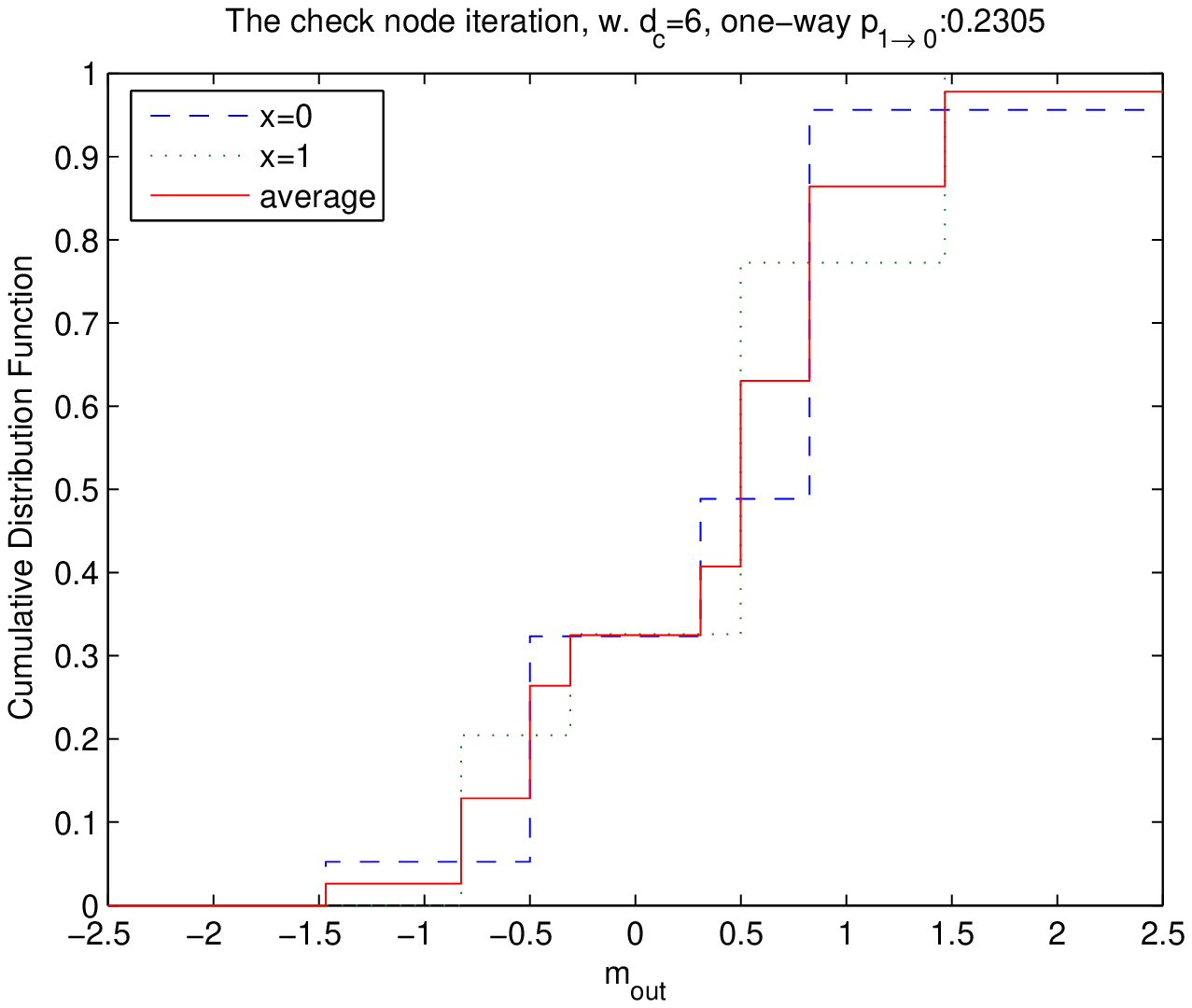}\\
\includegraphics[width=8.5cm,
keepaspectratio=true]{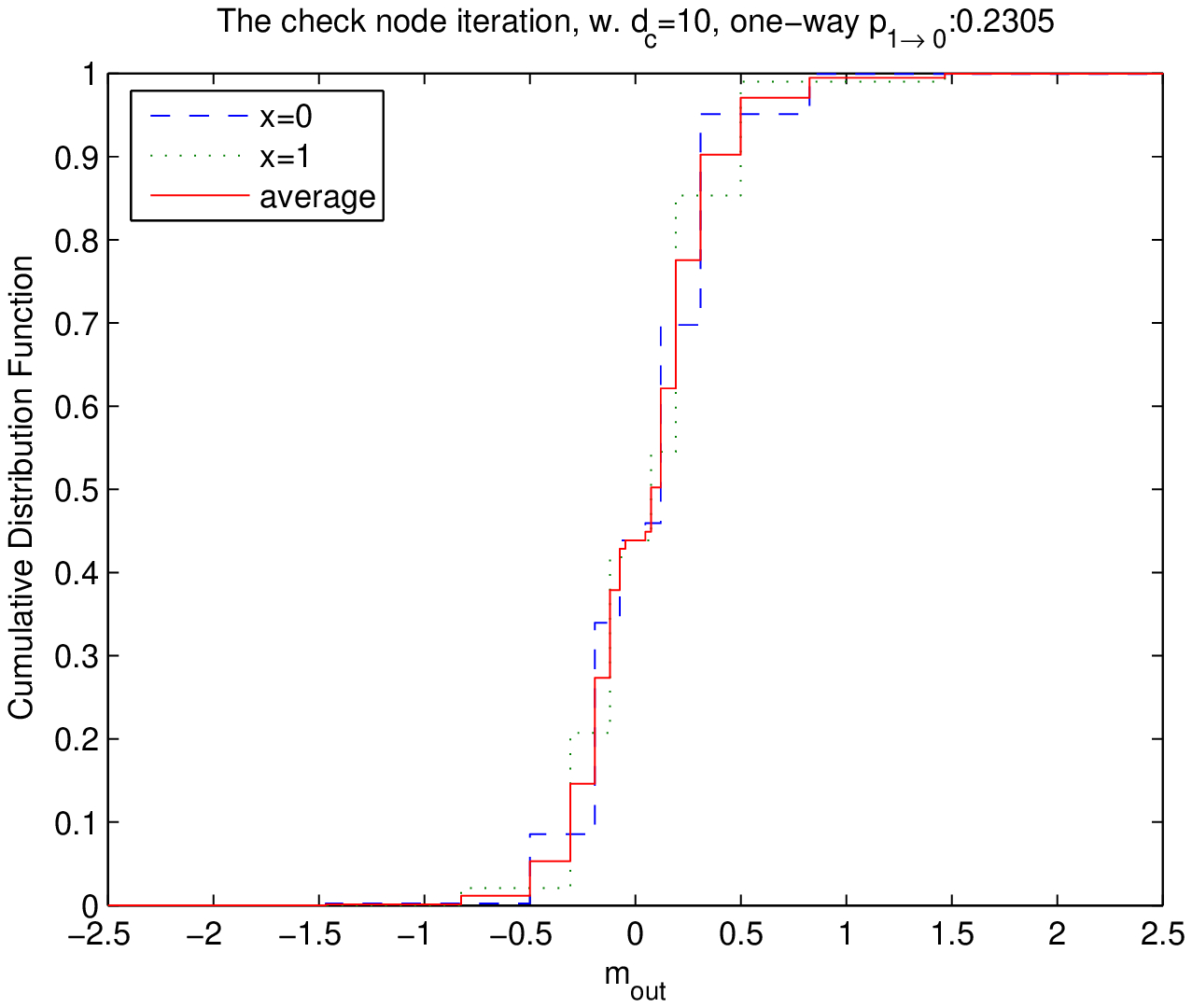}} \caption{Illustration of the weak
convergence of $Q^{(l-1)}_{a.p.}(0)$ and $Q^{(l-1)}_{a.p.}(1)$. One can see
that the convergence of $Q^{(l-1)}_{a.p.}(0)$ and $Q^{(l-1)}_{a.p.}(1)$ is
faster than the convergence of
$\frac{Q^{(l-1)}_{a.p.}(0)+Q^{(l-1)}_{a.p.}(1)}{2}$ and
$\delta_0$.\label{fig:weak-convergence}}
\end{figure}

\Figure{\ref{fig:weak-convergence}} illustrates the weak convergence predicted
by \Theorem{\ref{thm:typicality}} and depicts the convergence rates of
$Q^{(l-1)}_{a.p.}(0)\rightarrow Q^{(l-1)}_{a.p.}(1)$ and
$\frac{Q^{(l-1)}_{a.p.}(0)+Q^{(l-1)}_{a.p.}(1)}{2}\rightarrow\delta_0$.

Our typicality result can be viewed as a complementing theorem of
the concentration theorem in [\Corollary{2.2} of
\cite{KavcicMaMitzanmacher03}], where a constructive method of
finding a typical coset-defining syndrome is not specified.
Besides the theoretical importance, we are now on a solid basis to
interchangeably use the linear LDPC codes and the LDPC coset codes
when the check node degree is of moderate size. For instance, from
the implementation point of view, the hardware uniformity of
linear codes makes them a superior choice compared to any other
coset code. We can then use the fast density evolution
\cite{JinRichardson04} plus the coset code ensemble to optimize
the degree distribution for the linear LDPC codes. Or instead of
simulating the codeword-averaged performance of linear LDPC codes,
we can simulate the error probability of the all-zero codeword in
the coset code ensemble, in which the efficient LDPC encoder
\cite{RichardsonUrbanke01b} is not necessary.

\subsection{Revisiting the Belief Propagation Decoder\label{subsec:side-results-BP}}
Two known facts about the BP algorithm and the density evolution
method are as follows. First, the BP algorithm is optimal for any
cycle-free network, since it exploits the independence of the
incoming LLR message. Second, by the cycle-free convergence
theorem, the traditional density evolution is able to predict the
behavior of the BP algorithm (designed for the tree structure) for
$l_0$ iterations, even when we are focusing on a Tanner graph of a
finite-length LDPC code, which inevitably has many cycles. The
performance of BP, predicted by density evolution, is outstanding
so that we ``implicitly assume" that the BP (designed for the tree
structure) is optimal for the first $l_0$ iterations in terms of
minimizing the {\it codeword-averaged} bit error rate (BER).
Theoretically, to be able to minimize the codeword-averaged BER,
the optimal decision rule inevitably must exploit the global
knowledge about all possible codewords, which is, however, not
available to the BP decoder. A question of interest is whether BP
is indeed optimal for the first $l_0$ iterations? Namely, with
only local knowledge about possible codewords, whether BP has the
same performance as the optimal detector with the global
information about the entire codebook and unlimited computational
power when we are only interested in the first $l_0$ iterations?
The answer is a straightforward corollary to
\Theorem{\ref{thm:degree-of-freedom}}, the convergence to perfect
projection, which provides the missing link regarding the
optimality of BP when only local observations (on the ${\mathcal N
}^{2l})$ are available.

\begin{theorem}[Local Optimality of the BP Decoder]\label{thm:optimality-bp}
Fix $i, l_0\in \NN$. For sufficiently large codeword length~$n$,
almost all instances in the random code ensemble have the property
that
 the BP decoder for $x_i$ after $l_0$ iterations, $\hat{X}_{BP}({\mathbf
Y}^{l_0})$, coincides with the optimal MAP bit detector
$\hat{X}_{MAP,l_0}({\mathbf Y}^{l_0})$, where $l_0$ is a fixed
integer. The MAP bit detector $\hat{X}_{MAP,l_0}(\cdot)$ uses the
same number of observations as in $\hat{X}_{BP}(\cdot)$ but is
able to exploit the global knowledge about the entire codebook.
\end{theorem}
\begin{thmproof}{Proof:}
When the support tree ${\mathcal N}^{2l_0}_{(i,j)}$ is perfectly
projected, the local information about the tree-satisfying strings
is equivalent to the global information about the entire codebook.
Therefore, the extra information about the entire codebook does
not benefit the decision maker, and
$\hat{X}_{BP}(\cdot)=\hat{X}_{MAP,l_0}(\cdot)$.
\Theorem{\ref{thm:degree-of-freedom}} shows that ${\mathcal
N}^{2l_0}_{(i,j)}$ converges to perfect projection in probability,
which in turn implies that for sufficiently large $n$, BP decoder
is locally optimal for almost all instances of the code ensemble.
\end{thmproof}

{\it Note:} Even when limiting ourselves to symmetric memoryless
channels, this local optimality of BP can only be
proved\footnote{The existing cycle-free convergence theorem along
does not guarantee the local optimality of BP.} by the convergence
to perfect projection.  \Theorem{\ref{thm:optimality-bp}} can thus
be viewed as a completion of the classical density evolution for
symmetric memoryless channels.

\section{Conclusions\label{sec:conclusions}}
In this paper, we have developed a codeword-averaged density
evolution, which allows analysis of general {\it non-symmetric}
memoryless channels. An essential perfect projection convergence
theorem has been proved by a constraint propagation argument and
by analyzing the behavior of random matrices. With this perfect
projection convergence theorem, the theoretical foundation of the
codeword-averaged density evolution is well established. Most of
the properties of symmetric density evolution have been
generalized and proved for the codeword-averaged density evolution
on non-symmetric channels, including monotonicity, distribution
symmetry, and stability. Besides a necessary stability condition,
a sufficient stability condition has been stated with convergence
rate arguments and a simple proof.

The typicality of the linear LDPC code ensemble has been proved by
the weak convergence (w.r.t.\ $d_c$) of the evolved densities in
our codeword-averaged density evolution. Namely, when the check
node degree is sufficiently large (e.g.\ $d_c\geq 6$), the
performance of the linear LDPC code ensemble is very close to
(e.g.\ within $0.05\%$) the performance of the LDPC coset code
ensemble. One important corollary to the perfect projection
convergence theorem is the optimality of the belief propagation
algorithms when the global information about the entire codebook
is accessible. This can be viewed as a completion of the theory of
classical density evolution for symmetric memoryless channels.

 Extensive simulations have been presented, the degree distribution has been optimized for z-channels, and possible
applications of our results have been discussed as well. From both
practical and theoretical points of view,  our codeword-averaged
density evolution offers a straightforward and successful
generalization of the traditional symmetric density evolution for
general non-symmetric memoryless channels.

\appendices

\section{Proof of \Theorem{\ref{thm:degree-of-freedom}}\label{app:proof-convergence}}
We  first introduce the following corollary:
\begin{corollary}[Cycle-free Convergence]
For a sequence $l_n=\frac{4}{9}\frac{\ln
n}{\ln(d_v-1)+\ln(d_c-1)}$,   we have for any $i_0,j_0$,
    \begin{eqnarray}
        {\PP}\left(\mbox{${\mathcal
        N}^{2l_n}_{(i_0,j_0)}$ is
    cycle-free}\right)= 1-\bigorder\left(n^{-1/9}\right).\nonumber
    \end{eqnarray}\label{cor:cycle-free}
\end{corollary}

\begin{thmproof}{Proof of \Theorem{\ref{thm:degree-of-freedom}}:} In this proof,  the
subscript $(i_0,j_0)$ will be omitted  for notational simplicity.

We notice that if for any $l_n\geq l$, ${\mathcal
        N}^{2l_n}$ is  perfectly projected, then so
        is ${\mathcal
        N}^{2l}$.
    Choose $l_n=\frac{4}{9}\frac{\ln n}{\ln(d_v-1)+\ln(d_c-1)}$.
By \Corollary{\ref{cor:cycle-free}}, we have
\begin{eqnarray}
&&{\PP}(\mbox{${\mathcal
        N}^{2l}$ is perfectly projected})\nonumber\\
&&\geq{\PP}(\mbox{${\mathcal
        N}^{2l_n}$ is perfectly projected})\nonumber\\
&&\geq{\PP}\left(\left.\mbox{${\mathcal
        N}^{2l_n}$ is perfectly projected}\right|\mbox{${\mathcal
        N}^{2(l_n+1)}$ is cycle-free}\right)\nonumber\\
        &&~~~\cdot{\PP}\left(\mbox{${\mathcal
        N}^{2(l_n+1)}$ is cycle-free}\right)\nonumber\\
&&={\PP}\left(\left.\mbox{${\mathcal
        N}^{2l_n}$ is perfectly projected}\right|\mbox{${\mathcal
        N}^{2(l_n+1)}$ is cycle-free}\right)\nonumber\\
        &&~~~\cdot\left(1-\bigorder\left(n^{-1/9}\right)\right).\nonumber
\end{eqnarray}
We then need only to show
        that
\begin{eqnarray}
&&\hspace{-.7cm}{\PP}(\mbox{${\mathcal
        N}^{2l_n}$ is perfectly projected}|\mbox{${\mathcal
        N}^{2(l_n+1)}$ is cycle-free})\nonumber\\
        &=&1-\bigorder\left(n^{-0.1}\right).\label{eq:main-argument-for-full-rank-convergence}
\end{eqnarray}
To prove (\ref{eq:main-argument-for-full-rank-convergence}), we take a deeper
look at the incidence matrix (the parity check matrix) $\mathbf A$, and use the
$(3,5)$ regular code as our illustrative example. The proof is nonetheless
general for all regular code ensembles. Conditioning on the event that the
graph is cycle-free until depth $2\cdot 2$, we can transform $\mathbf A$ into
the form of (\ref{eq:big_eq}) by row and column swaps. Using  $\otimes$ to
denote the Kronecker product (whether it represents convolution or Kronecker
product should be clear from the context), (\ref{eq:big_eq}) can be further
expressed as follows.

\begin{figure*}[!t]
\normalsize
\setcounter{mytempeqncnt}{\value{equation}}
\setcounter{equation}{28}
\begin{equation}
{\mathbf A} \tiny= \tiny
\begin{pmat}[{....|.......................................|}]
 1&1&1&1&1& & & & & & & & & & & & & & & & & & & & & & & & & & & & & & & & & & & & & & & & &\cr\-
 1& & & & &1&1&1&1& & & & & & & & & & & & & & & & & & & & & & & & & & & & & & & & & & & & &\cr
 1& & & & & & & & &1&1&1&1& & & & & & & & & & & & & & & & & & & & & & & & & & & & & & & & &\cr
  &1& & & & & & & & & & & &1&1&1&1& & & & & & & & & & & & & & & & & & & & & & & & & & & & &\cr
  &1& & & & & & & & & & & & & & & &1&1&1&1& & & & & & & & & & & & & & & & & & & & & & & & &\cr
  & &1& & & & & & & & & & & & & & & & & & &1&1&1&1& & & & & & & & & & & & & & & & & & & & &\cr
  & &1& & & & & & & & & & & & & & & & & & & & & & &1&1&1&1& & & & & & & & & & & & & & & & &\cr
  & & &1& & & & & & & & & & & & & & & & & & & & & & & & & &1&1&1&1& & & & & & & & & & & & &\cr
  & & &1& & & & & & & & & & & & & & & & & & & & & & & & & & & & & &1&1&1&1& & & & & & & & &\cr
  & & & &1& & & & & & & & & & & & & & & & & & & & & & & & & & & & & & & & &1&1&1&1& & & & &\cr
  & & & &1& & & & & & & & & & & & & & & & & & & & & & & & & & & & & & & & & & & & &1&1&1&1&\cr\-
  & & & & &1& & & & & & & & & & & & & & & & & & & & & & & & & & & & & & & & & & & & & & & &\cdots\cr
\end{pmat}\normalsize\label{eq:big_eq}
\end{equation}
\setcounter{equation}{\value{mytempeqncnt}}

\addtocounter{equation}{1}

\hrulefill
\vspace*{4pt}
\end{figure*}

{\footnotesize
\begin{eqnarray}
{\mathbf A}~=\hspace{8cm}\nonumber\\  \left(\begin{array}{c|c|c}

{\mathbf I}_{1\times 1}\otimes(1,1,1,1,1)&{\mathbf 0}&{\mathbf 0}\\
\hline

{\mathbf I}_{5\times5}\otimes{1\choose1}&{\mathbf
I}_{10\times10}\otimes(1,1,1,1)&{\mathbf 0}\\
\hline

{\mathbf 0}&{\mathbf
I}_{40\times40}\otimes{1\choose1}&{\mathbf A}'_{80\times (n-45)}\\
\hline

{\mathbf 0}&{\mathbf 0}&{\mathbf A}''_{(\frac{3n}{5}-91)\times (n-45)}\\

\end{array}
 \right),\nonumber
\end{eqnarray}}\cmpt
where ${\mathbf I}_{a\times a}$ denotes the $a\times a$ identity matrix,
$\twobyone{{\mathbf A}'_{80\times (n-45)}}{{\mathbf
A}''_{(\frac{3n}{5}-91)\times (n-45)}}$ is the incidence matrix of the
equiprobable, bipartite subgraph, in which all $(n-45)$ variable nodes have
degree $d_v$, $80$ check nodes have degree $d_c-1$, and $(\frac{3n}{5}-91)$
check nodes have degree $d_c$. Conditioning on a more general event that the
graph is cycle free until depth $2(l_n+1)$ rather than $2\cdot 2$, we will have
{\tiny
\begin{eqnarray}
{{\mathbf A}} \tiny=\hspace{8.5cm}\nonumber\\\left(
\begin{array}{c|c|c|c}
\multicolumn{2}{c|}{
{\mathbf A}_{l_n}}&{\mathbf 0}&{\mathbf 0}\\
\hline

{\mathbf 0}&{\mathbf I}_{(5\cdot 8^{l_n-1})\times(5\cdot
8^{l_n-1})}\otimes{1\choose 1}&{\mathbf
I}_{(10\cdot 8^{l_n-1})\times(10\cdot 8^{l_n-1})}(1,1,1)&{\mathbf 0}\\
\hline

\multicolumn{2}{c|}{ {\mathbf 0}}&{\mathbf I}_{(5\cdot
8^{l_n})\times(5\cdot 8^{l_n})}\otimes{1\choose 1}&{\mathbf
A}'\\
\hline

\multicolumn{2}{c|}{ {\mathbf 0}}&{\mathbf 0}&{\mathbf A}''
\end{array}
\right),\nonumber
\end{eqnarray}}\cmpt
where ${\mathbf A}_{l_n}$ corresponds to the incidence matrix of
the cycle-free graph of depth $2l_n$. ${{\mathbf
A}'\choose{\mathbf A}''}$ is the incidence matrix with rows (check
nodes) in ${\mathbf A'}$ and ${\mathbf A}''$ having degree
$(d_c-1)$ and $d_c$. For convenience, we denote the blocks in
$\mathbf A$ as
\begin{eqnarray}
{\mathbf A}=\left(
\begin{array}{c|c|c|c}
\multicolumn{2}{c|}{
{\mathbf A}_{l_n}}&{\mathbf 0}&{\mathbf 0}\\
\hline

{\mathbf 0}&{\mathbf T}_{l_n}&{\mathbf
U}_{l_n+1}&{\mathbf 0}\\
\hline

\multicolumn{2}{c|}{ {\mathbf 0}}&{\mathbf T}_{l_n+1}&{\mathbf
A}'\\
\hline

\multicolumn{2}{c|}{ {\mathbf 0}}&{\mathbf 0}&{\mathbf A}''
\end{array}
\right).\nonumber
\end{eqnarray}

Then ${\mathcal  N}^{2l_n}$ is {\it not} perfectly projected if
and only if there exists a {\it non-zero} row vector $({\mathbf
r}|{\mathbf 0}|{\mathbf 0})$ such that
\begin{eqnarray}
({\mathbf r}|{\mathbf 0}|{\mathbf
0})\in\mbox{RowSpace}\left(\begin{array}{c|c|c|c}

{\mathbf 0}&{\mathbf T}_{l_n}&{\mathbf
U}_{l_n+1}&{\mathbf 0}\\
\hline

\multicolumn{2}{c|}{ {\mathbf 0}}&{\mathbf T}_{l_n+1}&{\mathbf
A}'\\
\hline

\multicolumn{2}{c|}{ {\mathbf 0}}&{\mathbf 0}&{\mathbf A}''

\end{array}\right),\label{eq:full-rank-equivalent-1}
\end{eqnarray}
and
\begin{eqnarray}
&&{\mathbf r}\mbox{ is not in the row space of ${\mathbf
A}_{l_n}$},\nonumber\\
&&\mbox{or equivalently $(\mathbf r|{\mathbf 0}|{\mathbf 0})$ is not in
RowSpace(${\mathbf A}_{l_n}|{\mathbf 0}|{\mathbf 0}$)}.\nonumber\\
\label{eq:full-rank-equivalent-2}
\end{eqnarray}
Eqs.~(\ref{eq:full-rank-equivalent-2}) and
(\ref{eq:full-rank-equivalent-1}) say that there exists a
constraint $\mathbf r$ on the variable nodes of ${\mathcal
N}^{2l_n}$, which is not from the linear combination of those
check node equations within ${\mathcal N}^{2l_n}$, but rather is
imposed by the parity check equations outside ${\mathcal
N}^{2l_n}$. It can be easily proved that if the matrix ${{\mathbf
A}' \choose {\mathbf A}''}$ is of full row rank, then no such
$\mathbf r$ exists and ${\mathcal N}^{2l_n}$ is perfectly
projected.\footnote{Unfortunately, ${{\mathbf A}' \choose {\mathbf
A}''}$ is {\it not} of full row rank. We can only show that with
sufficiently large $n$, the row rank of ${{\mathbf A}' \choose
{\mathbf A}''}$ converges to the number of rows minus one by
methods similar to those in \cite{MillerCohen03}. A simple
constraint propagation argument is still necessary for this
approach. } Instead of proving ${{\mathbf A}' \choose {\mathbf
A}''}$ is of full rank, we take a different approach, which takes
care of the constraint propagation.

From (\ref{eq:full-rank-equivalent-1}), we know that, for
$(\mathbf r|{\mathbf 0}|{\mathbf 0})$ to exist, there must exist a
{\it non-zero} row vector $({\mathbf 0}|{\mathbf s}|{\mathbf 0})$
such that
\begin{eqnarray}
({\mathbf 0}|{\mathbf s}|{\mathbf
0})\in\mbox{RowSpace}\left(\begin{array}{c|c|c|c}

\multicolumn{2}{c|}{ {\mathbf 0}}&{\mathbf T}_{l_n+1}&{\mathbf
A}'\\
\hline

\multicolumn{2}{c|}{ {\mathbf 0}}&{\mathbf 0}&{\mathbf A}''

\end{array}
\right),\label{eq:outside-constraints}
\end{eqnarray}
and
\begin{eqnarray}
{\mathbf s}&\in&\mbox{RowSpace}({\mathbf
U}_{l_n+1})\nonumber\\
&=&\mbox{RowSpace}\left(I_{(10\cdot 8^{l_n-1})\times(10\cdot
8^{l_n-1})}\otimes(1,1,1,1)\right).\nonumber\\ \label{eq:aligned-bits}
\end{eqnarray}
From (\ref{eq:aligned-bits}), the 1's in $\mathbf s$ must be
aligned such that four neighboring bits should have the same
value; for example, ${\mathbf
s}=(111100001111000000001111\cdots00001111)$.

Any non-zero $\mathbf s$ satisfying (\ref{eq:outside-constraints})
is generated by ${\mathbf T}_{l_n+1}$. By applying  the row
symmetry in ${\mathbf A}'$, we see that  the 1's in any  ${\mathbf
s}$ are uniformly distributed among all these $5\cdot 8^{l_n}$
bits. Therefore, conditioning on the event that there exists a
not-all-one $\mathbf s$ satisfying
Eq.~(\ref{eq:outside-constraints}), the probability that $\mathbf
s$ satisfies Eq.~(\ref{eq:aligned-bits}) is
\begin{eqnarray}
&&\hspace{-.7cm}\PP\left(\mbox{$\mathbf s$ satisfies
Eq.~(\ref{eq:aligned-bits})}|\mbox{$\exists\mathbf
s$ satisfies Eq.~(\ref{eq:outside-constraints}) and is not $\mathbf 1$}\right)\nonumber\\
&=&\PP\left(\mbox{the 1's in ${\mathbf s}$ are
aligned}|\right.\nonumber\\
&&\hspace{2cm}\left.\mbox{$\exists\mathbf s$ satisfies
Eq.~(\ref{eq:outside-constraints}) and is not $\mathbf 1$}\right)\nonumber\\
&=&\sum_{a=1}^{10\cdot 8^{l_n-1}-1}\frac{{10\cdot 8^{l_n-1}\choose a}}{{5\cdot
8^{l_n}\choose 4a}}\cdot\PP(\mbox{there are $4a$ ones in ${\mathbf
s}$})\nonumber\\
&\leq&\frac{{10\cdot 8^{l_n-1}\choose 1}}{{5\cdot 8^{l_n}\choose
4}}{\bigorder}\left(\left(\frac{1}{((d_v-1)(d_c-1))^{l_n}}\right)^{d_c-2}\right)\nonumber\\
&=&{\bigorder}\left(n^{-\frac{4}{9}(d_c-2)}\right).
\label{eq:probability-of-one-propagated-s}
\end{eqnarray}
 The last inequality follows from the assumption that
${\mathbf s}$ is neither all-zero nor all-one. The reason why we
can exclude the case that $\mathbf s$ is all-one is that, if $d_v$
is odd, then there is an even number of 1's in each column of
${\mathbf T}_{l_n}$. Since there is only one 1 in each column of
${\mathbf U}_{l_n+1}$, by (\ref{eq:full-rank-equivalent-1}), an
all-one $\mathbf s$ can only generate an all-zero ${\mathbf r}$,
which puts no constraints on ${\mathcal N}^{2l_n}_{(i_0,j_0)}$. If
$d_v$ is even, by the same reasoning, an all-one $\mathbf s$ will
generate ${\mathbf{r}}$ of the form $(00\cdots
0\overbrace{11\cdots 1}^{5\cdot 8^{l_n-1}})$.
 Nevertheless, when $d_v$ is
even, this specific type of $\mathbf r$ is in the row space of
${\mathbf A}_{l_n}$, which does not fulfill the requirement in
(\ref{eq:full-rank-equivalent-2}). From the above reasoning, we
can exclude the all-one $\mathbf s$.

Let $m_r$ denote the number of rows of ${{\mathbf
A}'\choose{\mathbf A}''}$ minus $\Rank({{\mathbf
A}'\choose{\mathbf A}''})$. The number of vectors $\mathbf s$
satisfying (\ref{eq:outside-constraints}) is upper bounded by
$2^{m_r}$. By (\ref{eq:probability-of-one-propagated-s}),
\Proposition{\ref{prop:expectation-of-mr}} (which will be formally
stated and proved later), and the union bound, we have
\begin{eqnarray}
&&\hspace{-.7cm}{\PP}(\mbox{${\mathcal
        N}^{2l_n}_{(i_0,j_0)}$ is not perfectly projected}|\mbox{${\mathcal
        N}^{2(l_n+1)}_{(i_0,j_0)}$ is cycle-free})\nonumber\\
&=&{\PP}(\mbox{$\exists{\mathbf r}$ satisfying
(\ref{eq:full-rank-equivalent-1}) and
(\ref{eq:full-rank-equivalent-2})})\nonumber\\
&=&{\PP}(\mbox{$\exists{\mathbf s}$, which satisfies
(\ref{eq:outside-constraints}) and
(\ref{eq:aligned-bits}), but is not all-one})\nonumber\\
&\leq& n^{1.1}\cdot\PP\left(\mbox{$\mathbf s$ satisfies
Eq.~(\ref{eq:aligned-bits})}|\right.\nonumber\\
&&\hspace{2cm}\left.\mbox{$\exists\mathbf s$ satisfies
Eq.~(\ref{eq:outside-constraints}) and is not $\mathbf
1$}\right)\nonumber\\
&&\cdot\PP\left(\mbox{\# of $\mathbf s$ is smaller than
$n^{1.1}$}\right)\nonumber\\
&&+ \PP\left(\mbox{\# of $\mathbf s$ is larger than
$n^{1.1}$}\right)\nonumber\\
&=&
n^{1.1}\bigorder(n^{-\frac{4}{9}(d_c-2)})+{\PP}(2^{m_r}>n^{1.1})\nonumber\\
&=&\bigorder(n^{-0.1}), ~~\forall d_c\geq 5.\label{eq:case-dc-larger-than-4}
\end{eqnarray}
 To prove the case
$d_c<5$, we focus on the probability that the constraints
propagate two levels rather than just one level, i.e.\ instead of
(\ref{eq:main-argument-for-full-rank-convergence}), we focus on
proving the following statement:
\begin{eqnarray}
&&\hspace{-.7cm}{\PP}(\mbox{${\mathcal
        N}^{2l_n}_{(i_0,j_0)}$ is perfectly projected}|\mbox{${\mathcal
        N}^{2(l_n+2)}_{(i_0,j_0)}$ is cycle-free})\nonumber\\
        &=&1-\bigorder\left(n^{-0.1}\right).\nonumber
\end{eqnarray}
Most of the analysis remains the same. The conditional probability
in (\ref{eq:probability-of-one-propagated-s}) will be replaced by
\begin{eqnarray}
&&\hspace{-.7cm}{\PP}(\mbox{$({\mathbf 0}|{\mathbf
s}|{\mathbf 0})$ is able to propagate two levels}|\mbox{$\exists\mathbf s$ satisfying (\ref{eq:outside-constraints})})\nonumber\\
&=&{\PP}(\mbox{$({\mathbf 0}|{\mathbf s}|{\mathbf 0})$ propagates the 2nd
level}|\nonumber\\
&&~~~~~\mbox{$({\mathbf 0}|{\mathbf s}|{\mathbf 0})$ propagates the 1st
level}, \mbox{$\exists\mathbf s$ satisfying (\ref{eq:outside-constraints})})\nonumber\\
&&\cdot{\PP}(\mbox{$({\mathbf 0}|{\mathbf
s}|{\mathbf 0})$ propagates the 1st level}|\mbox{$\exists\mathbf s$ satisfying (\ref{eq:outside-constraints})})\nonumber\\
&=&\sum_{a,b}\frac{{10\cdot 8^{l_n-1}\choose a}}{{5\cdot 8^{l_n}\choose
4a}}\frac{{10\cdot 8^{l_n}\choose
b}}{{5\cdot 8^{l_n+1}\choose 4b}}\nonumber\\
&&~\cdot\PP(\mbox{$4a$ and $4b$ 1's to propagate the 2nd and 1st levels} )\nonumber\\
&\stackrel{(a)}{\leq}&\frac{{10\cdot 8^{l_n-1}\choose 1}}{{5\cdot
8^{l_n}\choose 4}}\frac{{10\cdot 8^{l_n1}\choose 4}}{{5\cdot 8^{l_n+1}\choose
4\cdot 4}}\nonumber\\
&=&{\mathcal O}\left(n^{-\frac{4}{9}(d_c^2-2d_c)}\right),\nonumber
\end{eqnarray}
where the inequality marked (a) follows from an analysis of the
minimum number of bits required for the constraint propagation
similar to that for the single level case. By this stronger
inequality and a bounding inequality similar to that in
(\ref{eq:case-dc-larger-than-4}), we thus complete the proof of
the case $d_c\geq 3$ for all regular codes of practical interest.

\end{thmproof}

{\it Note:} This constraint propagation argument shows that the
convergence to a perfectly projected tree is very strong. Even for
codes with redundant check node equations (not of full row rank),
it is probabilistically hard for the external constraints to
propagate inside and impose on the variable nodes within
${\mathcal N}^{2l}$. This property is helpful when we consider
belief propagation decoding on the alternative graph
representation as in \cite{KumarMilenkovicPrakash05}.

We close this section by stating the proposition regarding $m_r$,
the number of linearly dependent rows in ${{\mathbf
A}'\choose{\mathbf A}''}$. The proof is left to
\Append{\ref{app:proof-of-rank-proposition}}.
\begin{proposition}\label{prop:expectation-of-mr}
Consider the semi-regular code ensemble ${\mathcal
C}^n_{m',m''}(d_v,d_c)$ generated by equiprobable edge permutation
on a bipartite graph with $n$ variable nodes of degree $d_v$, and
$m'$ and $m''$ check nodes with respective degrees $(d_c-1)$ and
$d_c$. The corresponding parity check matrix is ${\mathbf
A}={{\mathbf A}'\choose{\mathbf A}''}$. With $m_r$ denoting the
number of linearly dependent rows in ${\mathbf A}$, i.e.\
$m_r:=m'+m''-\Rank({\mathbf A})$, we have
    \begin{eqnarray}
    {\EE}\{2^{m_r}\}=\bigorder(n),\nonumber
    \end{eqnarray}
which automatically implies
${\PP}\left(2^{m_r}>n^{1+\alpha}\right)={\PP}\left(m_r>\frac{(1+\alpha)\ln
n}{\ln 2}\right)=\bigorder(n^{-\alpha})$, for any $\alpha>0$.
\end{proposition}
\begin{corollary}
Let $R$ denote the rate of a regular LDPC code ensemble ${\mathcal
C}^n(d_v,d_c)$, i.e., $R=\frac{n-\Rank({\mathbf A})}{n}$, where
$\mathbf A$ is the corresponding parity check matrix. Then $R$
converges to $(n-m)/n$ in $L^1$, i.e.
    \begin{eqnarray}
    \lim_{n\rightarrow\infty}\EE\left\{\left|R-\frac{n-m}{m}\right|\right\}=0.\nonumber
    \end{eqnarray}
\end{corollary}

\vspace{.3cm}
\begin{proof}
    It is obvious that $R\geq\frac{n-m}{n}$. To show that $\limsup_{n\rightarrow\infty}\EE\{R-\frac{n-m}{n}\}=0$, we let
    $m_1=0$ and rewrite    $R=\frac{n-\Rank({\mathbf A})}{n}=\frac{n-m}{n}+\frac{m_r}{n}$.
    By \Proposition{\ref{prop:expectation-of-mr}} and the fact that $\frac{m_r}{n}\leq 1$, we have
    $\lim_{n\rightarrow\infty}\EE\{\frac{m_r}{n}\}=0$. This completes the proof.
\end{proof}
A stronger version of the convergence of $R$ with respect to the
block length $n$ can be found in \cite{MillerCohen03}.

\section{Proof of
\Proposition{\ref{prop:expectation-of-mr}}\label{app:proof-of-rank-proposition}}

We finish the proof of \Proposition{\ref{prop:expectation-of-mr}}
by first stating the following lemma.
\begin{lemma}\label{lem:n-choose-i} For all $0<k\in\NN, 0<i< n\in \NN$, we have
\begin{eqnarray}
\frac{{n \choose i}}{{kn \choose ki}}\leq
\sqrt{k}e^{\frac{1}{6}}2^{-(k-1)nH_2(i/n)}.\nonumber
\end{eqnarray}
\end{lemma}
\begin{thmproof}{Proof:}
By Stirling's double inequality,
    \begin{eqnarray}
    \sqrt{2\pi}n^{(n+\frac{1}{2})}e^{(-n+\frac{1}{12n+1})}<n!<\sqrt{2\pi}n^{(n+\frac{1}{2})}e^{(-n+\frac{1}{12n})},\nonumber
    \end{eqnarray}
we can prove
    \begin{eqnarray}
    e^{-\frac{1}{6}}<\frac{{n\choose \theta
    n}}{\frac{1}{\sqrt{2\pi}}2^{nH_2(i/n)}\sqrt{\frac{n}{i(n-i)}}}<1,\nonumber
    \end{eqnarray}
which immediately leads to the desired inequality.
\end{thmproof}

\begin{thmproof}{Proof of \Proposition{\ref{prop:expectation-of-mr}}:}
By the definition of $m_r$, we have $2^{m_r}=(\mbox{total \# of
codewords})/2^{n-m}$, where $m=m'+m''$. Then
\begin{eqnarray}
\EE\left\{\frac{2^{m_r}}{n}\right\}\leq \frac{2}{n2^{n-m}}
+\sum_{i=1}^{n-1}\frac{\EE\left\{\mbox{\# of codewords of weight
$i$}\right\}}{n2^{n-m}}.\nonumber
\end{eqnarray}
Using the enumerating function as in \cite{BurshteinMiller04,MillerCohen03} and
define $g(x)$ as
\begin{eqnarray}
g(i,x):=\frac{\left(\frac{(1+x)^{d_c-1}+(1-x)^{d_c-1}}{2}\right)^{m'}\left(\frac{(1+x)^{d_c}+(1-x)^{d_c}}{2}\right)^{m''}}{x^{id_v}},\nonumber
\end{eqnarray}
 the above quantity can be further upper bounded as follows.
 {\small
\begin{eqnarray}
&&\hspace{-.7cm}\EE\left\{\frac{2^{m_r}}{n}\right\}\nonumber\\ &\leq&
\frac{2}{n2^{n-m}}+\sum_{i=1}^{n-1}\frac{\frac{{n\choose i }}{{nd_v\choose
id_v}}\inf_{x>0}g(i,x)}{n2^{n-m}}\nonumber\\
&\leq&
\frac{2}{n2^{n-m}}+\sqrt{d_v}e^{1/6}\sum_{i=1}^{n-1}\frac{1}{n}2^{-(d_v-1)nH_2(i/n)}\frac{\inf_{x>0}g(i,x)}{2^{n-m}}\nonumber\\
&\leq&
\frac{2}{n2^{n-m}}+\sqrt{d_v}e^{1/6}\sum_{i=1}^{n-1}\frac{1}{n}2^{-d_vnH_2(i/n)}\frac{\inf_{x>0}g(i,x)}{2^{-m}},\nonumber\\
\label{eq:sum-of-f}
\end{eqnarray}}\cmpt
where the second inequality follows from
\Lemma{\ref{lem:n-choose-i}} and the third inequality follows from
the fact that the binary entropy function $H_2(\cdot)$ is upper
bounded by 1.

By defining
\begin{eqnarray}
f_n(i,x):=2^{m-d_vnH_2(i/n)}g(i,x),\nonumber
\end{eqnarray}
the summation in (\ref{eq:sum-of-f}) is upper bounded\footnote{The
range of $i$ is expanded here from a discrete integer set to a
continuous interval. } by
\begin{eqnarray}
\max_{i\in[0,n]}\inf_{x>0}f_n(i,x)\leq
\inf_{x>0}\max_{i\in[0,n]}f_n(i,x)\leq
\max_{i\in[0,n]}f_n(i,1).\nonumber
\end{eqnarray}
By simple calculus, $\max_{i\in[0,n]}f_n(i,1)$ is attained when
$i=n/2$. Since $f_n(n/2,1)=1$, the summation in
(\ref{eq:sum-of-f}) is bounded by 1 for all $n$, and therefore
\begin{eqnarray}
\limsup_{n\rightarrow\infty}\EE\left\{\frac{2^{m_r}}{n}\right\}\leq
\sqrt{d_v}e^{1/6}.\nonumber
\end{eqnarray}
The proof is complete.

\end{thmproof}

\section{Proof of \Corollary{\ref{cor:typi-z}}\label{app:proof-typicality-z}}
We prove one direction that
\begin{eqnarray}
p^*_{1\rightarrow 0, linear}&:=&\sup\left\{p_{1\rightarrow 0}>0:
\lim_{l\rightarrow\infty}p_{e,linear}^{(l)}=0\right\}\nonumber\\
&>&\sup\left\{p_{1\rightarrow 0}>0:
\lim_{l\rightarrow\infty}p_{e,coset}^{(l)}=0\right\}-\epsilon\nonumber\\
&:=&p^*_{1\rightarrow 0, coset}-\epsilon.\nonumber
\end{eqnarray}
The other direction that $p^*_{1\rightarrow 0, coset}>p^*_{1\rightarrow 0,
linear}-\epsilon$ can be easily obtained by symmetry.

By definition, for any $\epsilon>0$, we can find a sufficiently
large $l_0<\infty$ such that for a z-channel with one-way
crossover probability $p_{1\rightarrow 0}:=p^*_{1\rightarrow 0,
coset}-\epsilon$, $P^{(l_0)}_{coset}$ is in the interior of the
stability region. We note that the stability region  depends only
on the Bhattacharyya noise parameter of $P^{(l_0)}_{coset}$, which
is a continuous function with respect to convergence in
distribution. Therefore, by \Theorem{\ref{thm:typicality}}, there
exists a $\Delta\in\NN$ such that $\left\langle
P^{(l_0)}\right\rangle$ is also in the stability region. By the
definition of the stability region, we have
$\lim_{l\rightarrow\infty}p_{e,linear}^{(l)}=0$, which implies
$p^*_{1\rightarrow 0, linear}\geq p_{1\rightarrow 0}$. The proof
is thus complete.

\section{The Convergence Rates of (\ref{eq:typicality-temp2}) and (\ref{eq:typicality-temp3})\label{app:conv-rate}}
For (\ref{eq:typicality-temp2}), we will consider the cases that $k=0$ and
$k=1$ separately. By the BASC decomposition argument, namely, all non-symmetric
channels can be decomposed as the probabilistic combination of many BASCs, we
can limit our attention to simple BASCs rather than general memoryless
non-symmetric channels. Suppose $P^{(l-1)}_{a.p.}(0)$ and $P^{(l-1)}_{a.p.}(1)$
correspond to a BASC with crossover probabilities $\epsilon_0$ and
$\epsilon_1$. Without loss of generality, we may assume $\epsilon_0+\epsilon_1<
1$ because of the previous assumption that $\forall x\in\GF(2),
P^{(l-1)}_{a.p.}(x)(m=0)=0$. We then have
\begin{eqnarray}
\Phi_{P'_0}(k,\frac{r}{\Delta})&=&(1-\epsilon_0)e^{i\frac{r}{\Delta}\ln\frac{1-\epsilon_0+\epsilon_1}{1-\epsilon_0-\epsilon_1}}
\nonumber\\
&&+(-1)^k\epsilon_0e^{i\frac{r}{\Delta}\ln\frac{1+\epsilon_0-\epsilon_1}{1-\epsilon_0-\epsilon_1}}\nonumber\\
\mbox{and~~}\Phi_{P'_1}(k,\frac{r}{\Delta})&=&(1-\epsilon_1)e^{i\frac{r}{\Delta}\ln\frac{1+\epsilon_0-\epsilon_1}{1-\epsilon_0-\epsilon_1}}
\nonumber\\
&&+(-1)^k\epsilon_1e^{i\frac{r}{\Delta}\ln\frac{1-\epsilon_0+\epsilon_1}{1-\epsilon_0-\epsilon_1}}.\nonumber
\end{eqnarray}

By Taylor's expansion, for $k=0$, (\ref{eq:typicality-temp2})
becomes
\begin{eqnarray}
        &&\hspace{-.7cm}2\left(\frac{\Phi_{P'_0}(0,\frac{r}{\Delta})-\Phi_{P'_1}(0,\frac{r}{\Delta})}{2}\right)^\Delta\nonumber\\
&=&
2\left(i\left(\frac{1-\epsilon_0-\epsilon_1}{2}\right)\left(\frac{r}{\Delta}\right)\ln\left(\frac{1-\epsilon_0+\epsilon_1}{1+\epsilon_0-\epsilon_1}\right)\right.\nonumber\\
&&\left.\hspace{4.4cm}+\bigorder\left(\left(\frac{r}{\Delta}\right)^2\right)\right)^\Delta,\nonumber
\end{eqnarray}
which converges to zero with convergence rate
$\bigorder\left(\bigorder(\Delta)^{-\Delta}\right)$. For $k=1$, we have
\begin{eqnarray}
        &&\hspace{-.7cm}2\left(\frac{\Phi_{P'_0}(1,\frac{r}{\Delta})-\Phi_{P'_1}(1,\frac{r}{\Delta})}{2}\right)^\Delta\nonumber\\
&=&
2\left(\left(\epsilon_1-\epsilon_0\right)+\frac{i}{2}\left(\frac{r}{\Delta}\right)\left(
(1-\epsilon_0+\epsilon_1)\ln\frac{1-\epsilon_0+\epsilon_1}{1-\epsilon_0-\epsilon_1}
\right.\right.\nonumber\\
&&\left.\hspace{3cm}-
(1+\epsilon_0-\epsilon_1)\ln\frac{1+\epsilon_0-\epsilon_1}{1-\epsilon_0-\epsilon_1}
\right)\nonumber\\
&&\left.~~~~+\bigorder\left(\left(\frac{r}{\Delta}\right)^2\right)\right)^\Delta,\nonumber
\end{eqnarray}
which converges to zero with convergence rate
$\bigorder(\const^\Delta)$, where $\const$ satisfies
$|\epsilon_1-\epsilon_0|<\const<1$. Since the convergence rate is
determined by the slower of the above two, we have proven that
(\ref{eq:typicality-temp2}) converges to zero with rate
$\bigorder(\const^\Delta)$ for some $\const<1$.

Consider (\ref{eq:typicality-temp3}). Since we assume that the
input is not perfect, we have $\max(\epsilon_0,\epsilon_1)>0$. For
$k=0$, by Taylor's expansion, we have
\begin{eqnarray}
        &&\hspace{-.7cm}\left(\frac{\Phi_{P'_0}(0,\frac{r}{\Delta})+\Phi_{P'_1}(0,\frac{r}{\Delta})}{2}\right)^\Delta\nonumber\\
&=& \left(1+\frac{i}{2}\left(\frac{r}{\Delta}\right)\left(
(1-\epsilon_0+\epsilon_1)\ln\frac{1-\epsilon_0+\epsilon_1}{1-\epsilon_0-\epsilon_1}\right.\right.\nonumber\\
&&\left.\hspace{3cm}+
(1+\epsilon_0-\epsilon_1)\ln\frac{1+\epsilon_0-\epsilon_1}{1-\epsilon_0-\epsilon_1}
\right)\nonumber\\
&&\left.~~~~+\bigorder\left(\left(\frac{r}{\Delta}\right)^2\right)\right)^\Delta,\nonumber
\end{eqnarray}
which converges to
\begin{eqnarray}
e^{i\left(\frac{r}{2}\right)\left(
(1-\epsilon_0+\epsilon_1)\ln\frac{1-\epsilon_0+\epsilon_1}{1-\epsilon_0-\epsilon_1}+
(1+\epsilon_0-\epsilon_1)\ln\frac{1+\epsilon_0-\epsilon_1}{1-\epsilon_0-\epsilon_1}
\right)}\nonumber
\end{eqnarray}
with rate $\bigorder\left(\Delta^{-1}\right)$. For $k=1$, we have
\begin{eqnarray}
        &&\hspace{-.7cm}\left(\frac{\Phi_{P'_0}(1,\frac{r}{\Delta})+\Phi_{P'_1}(1,\frac{r}{\Delta})}{2}\right)^\Delta\nonumber\\
&=&
\left(\left(1-\epsilon_0-\epsilon_1\right)\left(\frac{e^{i\frac{r}{\Delta}\ln\frac{1-\epsilon_0+\epsilon_1}{1-\epsilon_0-\epsilon_1}}
+e^{i\frac{r}{\Delta}\ln\frac{1+\epsilon_0-\epsilon_1}{1-\epsilon_0-\epsilon_1}}}{2}\right)\right)^\Delta,\nonumber
\end{eqnarray}
which converges to zero with rate
$\bigorder\left((1-\epsilon_0-\epsilon_1)^\Delta\right)$. Since
the overall convergence rate is the slower of the above two, we
have proven that the convergence rate is
$\bigorder\left(\Delta^{-1}\right)$.


%
%
\bibliography{chihw}
\bibliographystyle{IEEEtran}

\begin{biography}{Chih-Chun Wang} received the B.E. degree in E.E.\ from National Taiwan University, Taipei, Taiwan in 1999,
the M.S.\ degree in E.E., the Ph.D.\ degree in E.E. from Princeton University
in 2002 and 2005, respectively. He is currently working with Prof. Poor and
Prof. Kulkarni in a post-doc position at Princeton University. He worked in
COMTREND Corporation, Taipei, Taiwan, from 1999-2000, and spent the summer of
2004 with Flarion Technologies. He will join the ECE department of Purdue
University as an assistant professor in January 2006. His research interests
are in optimal control, information theory and coding theory, especially on the
performance analysis of iterative decoding for LDPC codes.
\end{biography}
\begin{biography}{Sanjeev R.\ Kulkarni}
(M'91, SM'96, F'04) received the B.S. in Mathematics, B.S.\ in
E.E., M.S.\ in Mathematics from Clarkson University in 1983, 1984,
and 1985, respectively, the M.S.\ degree in E.E.\ from Stanford
University in 1985, and the Ph.D.\ in E.E.\ from M.I.T.\ in 1991.
\\From 1985 to 1991 he was a Member of the Technical Staff at
M.I.T.\ Lincoln Laboratory working on the modelling and processing
of laser radar measurements. In the spring of 1986, he was a
part-time faculty at the University of Massachusetts, Boston.
Since 1991, he has been with Princeton University where he is
currently Professor of Electrical Engineering.
 He spent January 1996 as a research fellow
at the Australian National University, 1998 with Susquehanna
International Group, and summer 2001 with Flarion Technologies.

Prof.~Kulkarni received an ARO Young Investigator Award in 1992, an NSF Young
Investigator Award in 1994, and several teaching awards at Princeton
University.  He has served as an Associate Editor for the IEEE Transactions on
Information Theory.  Prof.~Kulkarni's research interests include statistical
pattern recognition, nonparametric estimation, learning and adaptive systems,
information theory, wireless networks, and image/video processing.
\end{biography}

\begin{biography}{H.~Vincent Poor}
 (S'72, M'77, SM'82, F'77) received the Ph.D.\ degree in EECS
from Princeton University in 1977.  From 1977 until 1990, he was on the faculty
of the University of Illinois at Urbana-Champaign. Since 1990 he has been on
the faculty at Princeton, where he is the George Van Ness Lothrop Professor in
Engineering. Dr.~Poor's research interests are in the areas of statistical
signal processing and its applications in wireless networks and related fields.
Among his publications in these areas is the recent book {\it Wireless
Networks: Multiuser Detection in Cross-Layer Design} (Springer: New York, NY,
2005).

Dr.~Poor is a member of the National Academy of Engineering and is a Fellow of
the American Academy of Arts and Sciences. He is also a Fellow of the Institute
of Mathematical Statistics, the Optical Society of America, and other
organizations.  In 1990, he served as President of the IEEE Information Theory
Society, and he is currently serving as the Editor-in-Chief of these {\it
Transactions}. Recent recognition of his work includes the Joint Paper Award of
the IEEE Communications and Information Theory Societies (2001), the NSF
Director's Award for Distinguished Teaching Scholars (2002), a Guggenheim
Fellowship (2002-03), and the IEEE Education Medal (2005).

\end{biography}


\end{document}